\documentclass[%
 reprint,
superscriptaddress,
 amsmath,amssymb,
 aps,
prb,
]{revtex4-2}

\usepackage{graphicx}
\usepackage{dcolumn}
\usepackage{bm}
\usepackage{braket}
\usepackage{hyperref,xcolor,siunitx}
\hypersetup{
	colorlinks,
	linkcolor={magenta},
	citecolor={magenta},
	urlcolor={blue!70!black}
}

\begin{document}

\title{Topological Lasing from Thouless Pumping in Bilayer Photonic Crystal}
\author{D.-H.-Minh Nguyen}
\email{d.h.minh.ng@gmail.com}
\affiliation{Donostia International Physics Center, 20018 Donostia-San Sebasti\'an, Spain}
\affiliation{Advanced Polymers and Materials: Physics, Chemistry and Technology, Chemistry Faculty (UPV/EHU), Paseo M. Lardizabal 3, 20018 San Sebastian, Spain}
\author{Dung Xuan Nguyen}
\email{dungmuop@gmail.com}
\affiliation{Brown Theoretical Physics Center and Department of Physics, Brown University, 182 Hope Street, Providence, Rhode Island 02912, USA}
\affiliation{Center for Theoretical Physics of Complex Systems, Institute for Basic Science (IBS), Daejeon, 34126, Republic of Korea}
\author{H. Chau Nguyen}
\affiliation{Naturwissenschaftlich-Technische Fakult\"at, Universit\"at Siegen, Walter-Flex-Stra{\ss}e 3, 57068 Siegen, Germany}
\author{Thibaud Louvet}
\affiliation{Ecole Centrale de Lyon, INSA Lyon, Universit\'e  Claude Bernard Lyon 1, CPE Lyon, CNRS, INL, UMR5270, Ecully 69130, France}
\author{Emmanuel Drouard}
\affiliation{Ecole Centrale de Lyon, INSA Lyon, Universit\'e  Claude Bernard Lyon 1, CPE Lyon, CNRS, INL, UMR5270, Ecully 69130, France}
\author{Xavier Letartre}
\affiliation{Ecole Centrale de Lyon, INSA Lyon, Universit\'e  Claude Bernard Lyon 1, CPE Lyon, CNRS, INL, UMR5270, Ecully 69130, France}
\author{Dario Bercioux}
\email{dario.bercioux@dipc.org}
\affiliation{Donostia International Physics Center, 20018 Donostia-San Sebasti\'an, Spain}
\affiliation{IKERBASQUE, Basque Foundation for Science, Euskadi Plaza, 5, 48009 Bilbao, Spain}
\author{Hai Son Nguyen}
\email{hai-son.nguyen@ec-lyon.fr}
\affiliation{Ecole Centrale de Lyon, INSA Lyon, Universit\'e  Claude Bernard Lyon 1, CPE Lyon, CNRS, INL, UMR5270, Ecully 69130, France}
\affiliation{IUF, Universi\'e de France, Paris 75231, France}

\date{\today}

\begin{abstract}
Topological lasing leverages concepts from topological physics to achieve single-mode light amplification within topological bandgaps, offering robustness against fabrication imperfections. Recent advances in microelectromechanical systems (MEMSs) and phase-change materials (PCMs) at the subwavelength scale promise new avenues for dynamically reconfigurable topological lasers, enabling robust and tunable nanoscale light sources.  Here, we numerically demonstrate a dynamically reconfigurable lasing action at telecom wavelengths in a bilayer photonic crystal through the mechanisms of Thouless pumping. By designing two competing periodic potentials -- one slowly translating photonic grating atop another stationary one -- we observe a transition between a topological pumping regime and conventional mode oscillation. A carefully engineered heterojunction between these phases supports a robust lasing mode that can be dynamically tuned via MEMSs or reversible PCMs. Our work establishes bilayer photonic crystals as a programmable platform for achieving topological light sources, showcasing a potential pathway for merging topological photonics with reconfigurable photonic devices.
\end{abstract}
\maketitle
\date{\today}

\let\oldaddcontentsline\addcontentsline
\renewcommand{\addcontentsline}[3]{}

\section{Introduction}
Topological photonics offers powerful tools to engineer novel states of light, using topological protection to create robust optoelectronic devices~\cite{Lu2014,Ota2018,Ozawa2019,Smirnova2020rev,Price2022,Szameit2024}. A prominent example is topological lasing, where light amplification occurs in edge states immune to disorder, offering enhanced robustness against fabrication imperfections. These systems typically rely on edge modes, such as those from topological insulators~\cite{Harari2018,Bandres2018,Choi2021}, Jackiw-Rebbi interface states~\cite{Haus1976,Sekartedjo1984,Yang2022,Tian2023perovskite,Ma2023room,Scherrer2024single}, or valley-Hall effects~\cite{Zhong2020,Zeng2020,Gong2020,Smirnova2020,Noh2020experimental,Duan2023valley,Hong2025three}. However, most  demonstrations to date have focused on static platforms with fixed geometries, lacking reconfigurability -- a key limitation for applications requiring adaptive or programmable functionalities~\cite{Gu2022,Zhao2024,Li2024}. Meanwhile, another cornerstone of topological physics is the model of Thouless pumping, which presents quantized transport driven by a slowly varying potential~\cite{Thouless1983}, has never been considered in this perspective. Although Thouless pumping has been demonstrated in photonic systems~\cite{Wang2022,Citro2023,Yang2024,Sun2024,Song2024,Ravets2024} --- most notably in coupled waveguide arrays --- its integration into practical devices like lasers or LEDs remains elusive.

Recent advances in bilayer photonic crystal slabs have opened new opportunities for light control, enabling phenomena such as moiré photonics~\cite{Tang2023,saadi2025}, chiral responses~\cite{Han2022,Gromyko2024}, optical singularities~\cite{Ni2024,Zhou2024}, asymmetric radiating metasurfaces~\cite{Lee2023,Zhuang2024,zhuang2025}, and synthetic momenta for topological physics~\cite{Lee2022,Minh2023}. Crucially, the ability to dynamically shift one layer relative to the other using microelectromechanical systems (MEMS) offers a unique mechanism to tailor band structures and realize tunable photonic functionalities~\cite{Tang2024hBN,Tang2025}. An alternative approach for achieving reconfigurability is through the integration of phase-change materials (PCMs)~\cite{Li2024}, such as germanium-antimony-tellurium (GST) or antimony trisulfide ($\text{Sb}_2\text{S}_3$). These materials exhibit reversible transitions between amorphous and crystalline phases, accompanied by pronounced changes in their optical properties. Incorporating PCMs into photonic structures enables dynamic control of light–matter interactions, making possible non-volatile switching, reconfigurable metasurfaces, and even programmable topological phases.

Here, inspired by a Thouless pumping model in a bipartite potential, we demonstrate the tunability of a topological lasing mode by designing a heterojunction configuration incorporating PCMs. This heterojunction supports a high-Q topological interface mode, which can be dynamically tuned using MEMSs or switched by inducing the crystal-to-amorphous transition in the PCM. Our numerical findings extend Thouless’s fundamental concept into a versatile photonic platform, providing a blueprint for engineering photonic devices that exploit topological robustness and tunability for advanced light manipulation and lasing applications.
\section{Results}
\subsection{Conceptual Overview}
Thouless pumping can be illustrated by considering spinless particles in a one-dimensional (1D) space subject to a time- and space-dependent periodic potential $\mathcal{U}(x,t) = \mathcal{U}(x+\Lambda,t) = \mathcal{U}(x,t+T)$, where $\Lambda$ and $T$ are the periodicity in space and time, respectively. There are $N_\text{p}$ particles in each minimum of the potential. For simplicity, we assume $N_\text{p}=1$ and that each particle is in its ground state inside a well. As this potential is adiabatically shifted along the $x$-axis, the wells carry the particles with them, resulting in a displacement $\Lambda$ of the particles per time period $T$. Consequently, the integral of the current over a time period is an integer value, i.e., 1, giving rise to quantized transport. The adiabatic condition is met if the potential changes slowly enough that the particles remain in their ground state throughout the process.
\begin{figure}
	\centering
	\includegraphics[width=\linewidth]{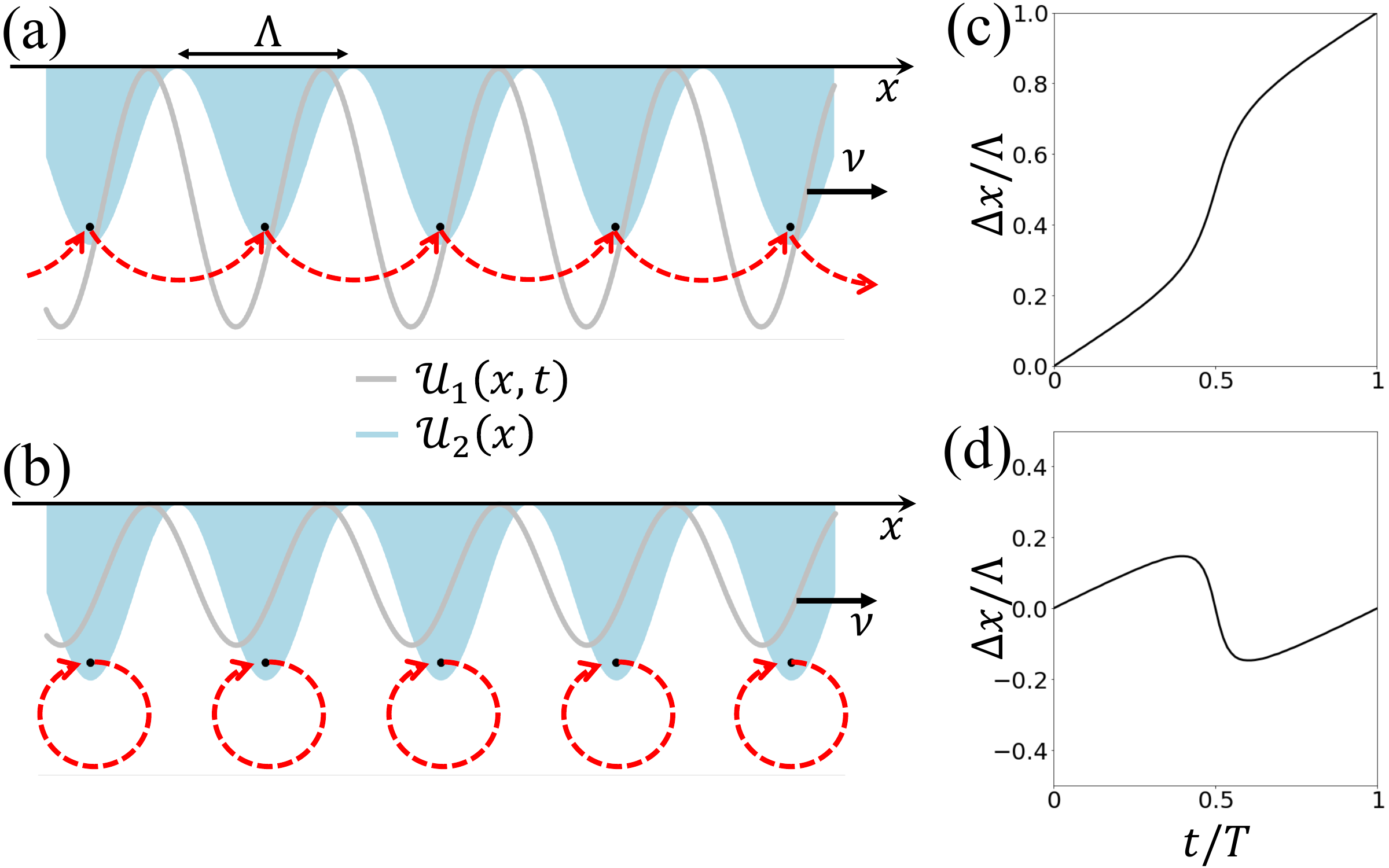}
	\caption{\label{fig:thouless} \textbf{Pumping versus Trapping}. (a), (b) Schematic drawings of two periodic potentials of the same periodicity where $\mathcal{U}_1(x,t)$ moves slowly and $\mathcal{U}_2(x)$ is stationary. The particles (denoted by black dots) can either be transported by $\mathcal{U}_1(x,t)$ to the next unit cell (a), or be pulled back by $\mathcal{U}_2(x)$ to their original positions (b). (c), (d) Exemplary changes in position of a particle in each case during a pumping cycle with sine potentials $\mathcal{U}_1(x,t) = 2.2\sin(2\pi x/\Lambda - 2\pi\nu t/\Lambda)$ (c) [$\mathcal{U}_1(x,t) = 1.2\sin(2\pi x/\Lambda - 2\pi\nu t/\Lambda)$ (d)] and $\mathcal{U}_2(x) = 1.5\sin(2\pi x/\Lambda)$.}
\end{figure}

In his seminal work~\cite{Thouless1983}, Thouless proposed an example of $\mathcal{U}(x,t)$ as a superposition of two periodic potentials: $\mathcal{U}(x,t) = \mathcal{U}_1(x,t) + \mathcal{U}_2(x)$. Here, $\mathcal{U}_1(x,t)=\mathcal{U}_1(x-\nu t)$ is a potential sliding slowly at a velocity $\nu$ whereas $\mathcal{U}_2(x)$ remains stationary. Both potentials share the same spatial period $\Lambda$, as depicted in Figs.~\ref{fig:thouless}({a}) and \ref{fig:thouless}({b}). Qian Niu later showed that the electrons in filled bands are locked into the stronger component of the bipartite potential~\cite{Niu1986}.
Specifically, the particles' motion depends on the competition between the mobile $\mathcal{U}_1(x,t)$ and the stationary $\mathcal{U}_2(x)$. While $\mathcal{U}_1(x,t)$ tends to push the particles to induce pumping, $\mathcal{U}_2(x)$ exerts a counteracting force and tends to localize them. When the driving potential $\mathcal{U}_1(x,t)$ dominates, the particles are transported by a distance $\Lambda$ at the end of a pumping cycle, as depicted in Figs.~\ref{fig:thouless}({a}) and~\ref{fig:thouless}({c}). Conversely, if the stationary potential is stronger [Fig.~\ref{fig:thouless}({b})], the particles return to their original positions at $t=T=\Lambda/\nu$ -- see Fig.~\ref{fig:thouless}({d}). We term the latter regime ``trapping'' of particles. Readers may refer to the Supplemental Video~1 for a dynamic visualization of Figs.~\ref{fig:thouless}({a}) and \ref{fig:thouless}({b}).
\subsection{Bilayer Photonic Crystal}
\begin{figure}
	\centering
	\includegraphics[width=\linewidth]{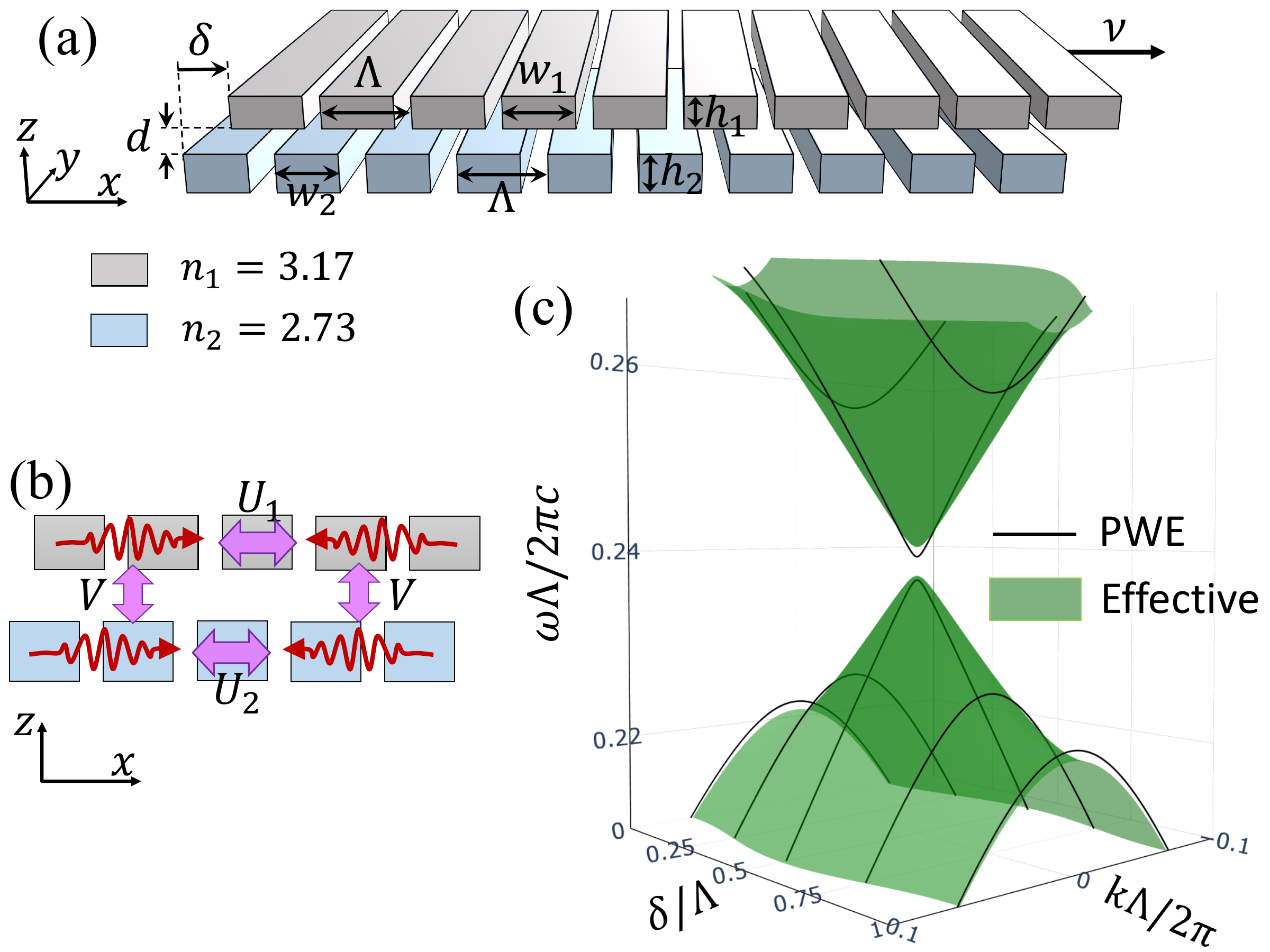}
	\caption{\label{fig:structure} \textbf{Bilayer photonic crystal}. (a) A bilayer photonic crystal composed of two gratings corrugated along the $x$ direction with the same period $\Lambda$. The gratings are made of materials with refractive indices $n_1$ and $n_2$. The upper layer slides slowly at velocity $\nu$, resulting in a lateral displacement $\delta=\nu t$ relative to the lower layer at time $t$. (b) Sketch of guided plane waves (red arrows) and their intra- and interlayer coupling mechanisms. (c) Dispersion of the two lowest guided TE bands of config.~(1) for various values of $\delta$. The green surfaces represent the effective model, while the black lines correspond to PWE simulations. Parameters: $w_1=0.8\Lambda$, $h_1=0.3\Lambda$, $w_2=0.81\Lambda$, $h_2=0.46\Lambda$, and $d=0.1\Lambda$.}
\end{figure}

We propose a realization of the bipartite potential described above in a 1D bilayer photonic crystal that comprises two parallel, high-contrast gratings separated by a distance $d$, as depicted in Fig.~\ref{fig:structure}({a}). The two gratings share the same period $\Lambda$ and are laterally displaced by $\delta$. 
They have thicknesses $h_\ell$, widths $w_\ell$, and refractive indices $n_\ell$, where $\ell=1,2$. All geometrical parameters are of subwavelength scale. As we shall see later, the upper and lower layers represent the potentials $\mathcal{U}_1(x,t)$ and $\mathcal{U}_2(x)$, respectively.

We focus on the transverse electric (TE) modes, whose $E_y$-components can be described by four guided plane waves: two traveling right and two traveling left, within the upper and lower layers -- see Fig.~\ref{fig:structure}({b}). In each layer, diffraction causes counter-propagating modes traveling with velocities $\pm v_\ell$ to couple at wave vector $k_x=\pi/\Lambda$ (i.e., the $X$ point of the first Brillouin zone) with a strength $U_\ell$ that depends on the grating's geometry and material. Plane waves traveling in the same direction but in different layers interact through their evanescent fields with strength~$V$. This coupling amplitude can be adjusted by varying the interlayer separation $d$. The effective Hamiltonian describing the four lowest-frequency guided modes reads~\cite{Nguyen:2018,Nguyen:2021a,Letartre2022}

\begin{equation}
	\label{eq:Hamiltonian}
	H(k,\delta) = \begin{pmatrix} 
		\omega_1+v_1k & U_1e^{ -i 2\pi \frac{\delta}{\Lambda}} & V &0\\
		U_1e^{ i 2\pi \frac{\delta}{\Lambda}} & \omega_1-v_1k &0 &V\\
		V &0 & \omega_2+v_2k & U_2 \\
		0& V & U_2 & \omega_2-v_2k	
	\end{pmatrix},
\end{equation}
where $k=k_x+\pi/\Lambda$ is the wave vector measured from the $X$ point; $\omega_1$ and $\omega_2$ are the frequency offsets in the upper and lower gratings. The phase $e^{\pm i2\pi\frac{\delta}{\Lambda}}$ in the intralayer coupling of the upper layer is induced by its displacement~$\delta$ -- see the Supplemental Information (SI) for a detailed derivation of the effective Hamiltonian.

We are interested in the two lowest TE-guided modes. Computing the Chern number of the lower gap separating these two modes in some typical cases, we find that the two cases $U_1>U_2$ and $U_1<U_2$ correspond to the pumping and trapping regimes in the bilayer photonic crystal, respectively. Although numerous designs of the bilayer crystal would yield similar results, we choose two exemplary configurations: config.~(1) with $(w_1, h_1, w_2, h_2) = (0.8, 0.3, 0.81, 0.46)\Lambda$, and config.~(2) with $(w_1, h_1, w_2, h_2) = (0.8, 0.3, 0.77, 0.5)\Lambda$. The former configuration has $U_1>U_2$, and the latter has $U_1<U_2$. The interlayer distance is fixed at $d=0.1\Lambda$. In both cases, the upper and lower gratings are made of materials with refractive indices $n_1=3.17$ and $n_2=2.73$, respectively. Potential materials for $n_1$ are amorphous silicon (a-Si), indium phosphide (InP), and gallium arsenide (GaAs), while the refractive index $n_2$ can be realized in titanium dioxide ($\text{Ti}\text{O}_2$), amorphous antimony trisulfide ($\text{Sb}_2\text{S}_3$) or aluminum arsenide (AlAs).

For the bilayer photonic crystal to experience all possible displacement configurations, we slowly translate the upper layer along the $x$-axis at a velocity $\nu\ll v_\ell$ so that $\delta=\nu t$ [Fig.~\ref{fig:structure}({a})]. The adiabatic condition is ensured since we choose the velocity to be sufficiently small so that the photonic modes at each instant are still accurately described by Hamiltonian~\eqref{eq:Hamiltonian}, with negligible influence from the motion. Owing to the lattice translation symmetry, the system is invariant under the transformation $\delta\rightarrow\delta+\Lambda$. The black lines in Fig.~\ref{fig:structure}({c}) show the two lowest bands of config.~(1) for several values of $\delta$, simulated using the plane wave expansion (PWE) method with the MIT Photonic Bands package~\cite{MPB}. By fitting the effective model's dispersion to the PWE simulation results at $\delta=0$, we retrieve all necessary parameters (see SI for details), and then plot the two lowest bands of Hamiltonian~\eqref{eq:Hamiltonian} in Fig.~\ref{fig:structure}({c}) (green surfaces), which shows good agreement between the effective model and PWE simulations. The two bands remain separated for all values of $\delta$. The lowest band reaches a maximum while the other minimizes at $k=0$ and $\delta=0.5\Lambda$.
\subsection{Photonic Pumping and Trapping}
We now show that the bilayer photonic crystal realizes the pumping and trapping regimes, with the mobile upper layer representing $\mathcal{U}_1(x,t)$ and the stationary lower layer corresponding to $\mathcal{U}_2(x)$. However, since we cannot define any ``particle'' in such a photonic crystal, what is pumped or trapped here is the electromagnetic field localized in the dielectric rods of this crystal, i.e., energy.
To track the motion of this localized field within a unit cell during a pumping cycle, we choose its center to be the Wannier center~\cite{Asboth2016,BlancodePaz2022}. The change in position of the field's center is given by
\begin{equation}
	\Delta x_\text{W}(t) = -\frac{\Lambda}{2\pi}\int_0^{t}dt'\int_{-\pi/\Lambda}^{+\pi/\Lambda}dk\Omega(k,t'),
\end{equation}
where $\Omega(k,t)=i\left(\braket{\frac{\partial u_1}{\partial k}|\frac{\partial u_1}{\partial t}} - \braket{\frac{\partial u_1}{\partial t}|\frac{\partial u_1}{\partial k}}\right)$ is the Berry curvature, and $\ket{u_1}$ represents the eigenstate of the lowest-frequency band. At the end of the pumping cycle, $x_\text{W}$ changes by $\Lambda$ if the localized field is successfully pumped, and remains unchanged (i.e., $\Delta x_\text{W}=0$) if it is trapped.
\begin{figure*}
	\includegraphics[width=\linewidth]{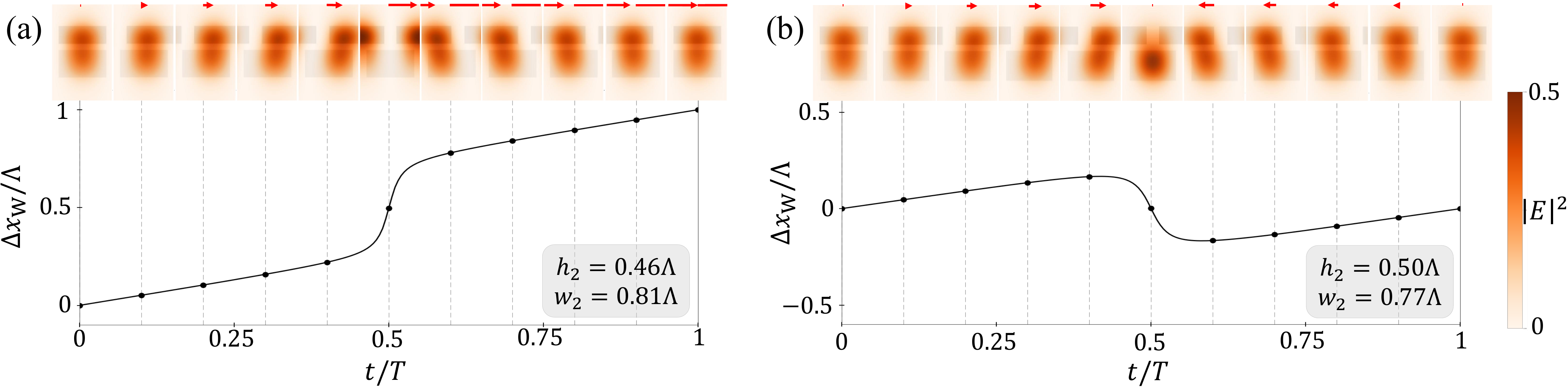}
	\caption{\label{fig:photonic thouless} \textbf{Photonic pumping and trapping}. (a), (b) Pumping and trapping of localized electric field in bilayer photonic crystal. The position of the localized field is denoted by the variation of the Wannier center, $\Delta x_\text{W}$, within a unit cell during a pumping period $T=\Lambda/\nu$. If $\Delta x_\text{W}$ changes by $\Lambda$ (a) or remains unchanged (b) at the end of the period, the field's center moves to the next unit cell or returns to its original position, respectively. The upper insets show the electric field distribution within a unit cell of the lowest mode at $k=0$ at various moments, obtained from PWE simulations. The red arrows indicate the corresponding displacement of the field. Here, the upper layer of the photonic crystal has $h_1=0.3\Lambda$ and $w_1=0.8\Lambda$ in both cases (a) and (b), and the parameters of the lower layer are shown in the gray insets. The color bar indicates the modulus squared of the normalized electric field~\cite{Joannopoulos2011}.}
\end{figure*}

The change in position of the Wannier center for both configurations (1) and (2) is efficiently computed using the effective model and visualized in Fig.~\ref{fig:photonic thouless}. Indeed, we see that the Wannier center shifts to the next unit cell in config.~(1), signifying pumping [Fig.~\ref{fig:photonic thouless}({a})], while it returns to its initial position at $t=T$ in config.~(2), indicating trapping [Fig.~\ref{fig:photonic thouless}({b})]. We look at the electric field distribution of the lowest mode obtained from PWE simulations to gain a deeper insight into how the fields vary. In both configurations, the electric field localizes in the dielectric rods of both layers across a wide range of wave number $k$, with the strongest localization at $k=0$ -- see the insets of Figs.~\ref{fig:photonic thouless}({a}) and \ref{fig:photonic thouless}({b}). However, different behaviors emerge: In config.~(1), the field follows the upper layer and gets dragged to the next unit cell, analogous to a particle pulled by the potential $\mathcal{U}_1(x,t)$. In config.~(2), the localized field tends to stay in the initial unit cell with the lower layer, despite being constantly driven by the upper layer. The electric field's behavior resembles that of a particle perturbed by $\mathcal{U}_1(x,t)$ but constantly pulled back by $\mathcal{U}_2(x)$. Therefore, depending on the configuration, the bilayer photonic crystal can emulate either Thouless pumping or trapping of electromagnetic field -- see the Supplemental Video~2 for animations of the insets in Fig.~\ref{fig:photonic thouless}.
\subsection{Reconfigurable Interface Mode}
\begin{figure*}
	\includegraphics[width=\linewidth]{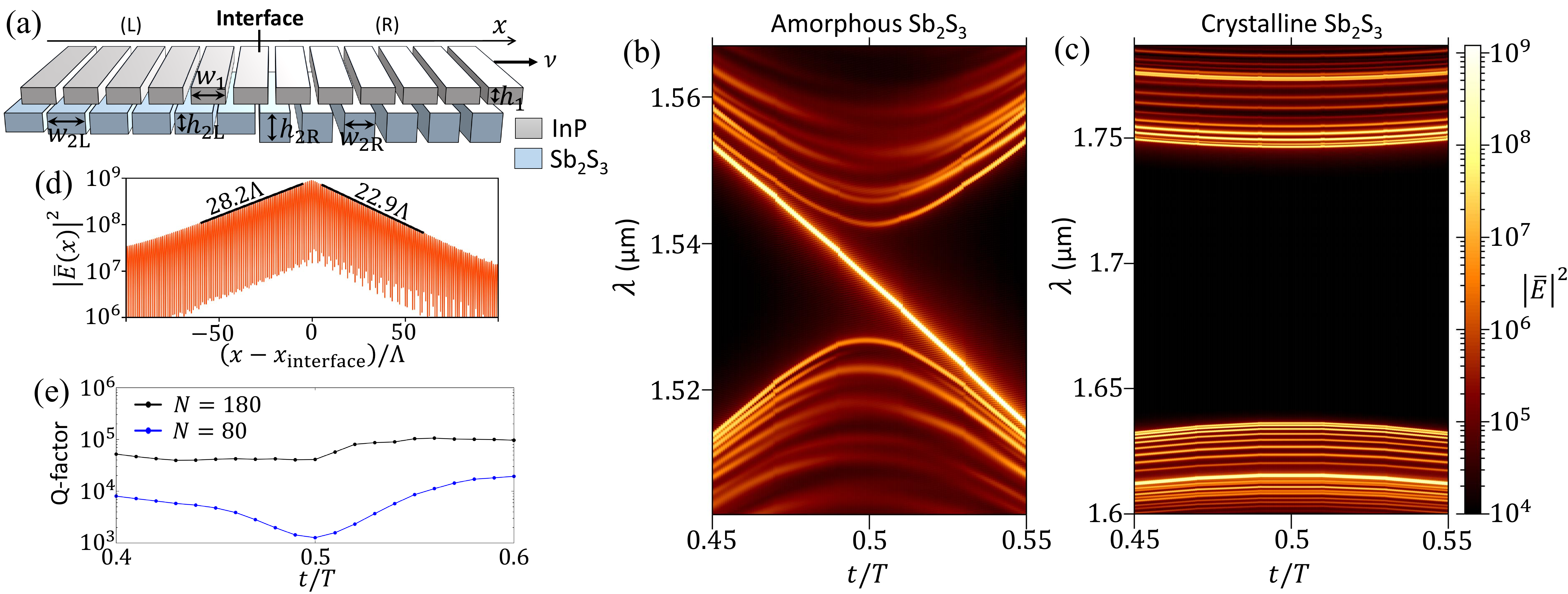}
	\caption{\label{fig:edge} \textbf{Heterojunction of bilayer photonic crystal.} (a) A photonic heterojunction in which the homogeneous upper grating slides with velocity $\nu$ while the heterogeneous $\text{Sb}_2\text{S}_3$ grating is stationary with heights and widths different on either side. The upper layer has $w_1=0.8\Lambda$ and $h_1=0.3\Lambda$. The left (right) side of the $\text{Sb}_2\text{S}_3$ layer has $w_{2\text{L}(\text{R})}=0.81(0.77)\Lambda$ and $h_{2\text{L}(\text{R})}=0.46(0.5)\Lambda$. The interlayer distance is $d=0.1\Lambda$. (b), (c) The spectra of TE modes in the heterojunction when $\text{Sb}_2\text{S}_3$ is in amorphous (b) and crystalline (c) phases, simulated by FDTD method with the number of periods on each side being $N=200$. (d) The electric field profile $|\bar{E}(x)|^2$ of the interface mode in (b) at $t=0.5T$, integrated over the $z$ direction. (e) The quality factor of the interface mode in (b) with respect to time for two heterojunction sizes: $N=180$ and $N=80$. The lattice period is $\Lambda=366$~nm.}
\end{figure*}

In the context of Thouless pumping, by considering time ($t$) as an additional dimension, the system can be examined in a $(1+1)$-D parameter space. Within this framework, the energy bands of the lattice are characterized by an invariant known as the Chern number~\cite{Asboth2016,Citro2023}, $C_n=-\Delta x_{\text{W}n}(T)/\Lambda$ with $n$ the band index, that identifies the topological phase of the system.
In our case, the pumping and trapping regimes have different Chern numbers for the lowest band: $C=-1$ and $C=0$, respectively, indicating two distinct topological phases in the $(1+1)$-D space. By constructing a heterojunction of two photonic crystals with different topological phases, we expect to observe a robust interface mode protected by the topological phase transition across this heterojunction. This interface mode is pumped through the frequency gap as the lattice varies in time~\cite{Asboth2016}. Hence, we consider a photonic heterojunction composed of config.~(1) on the left (L) and config.~(2) on the right (R), as illustrated in Fig.~\ref{fig:edge}({a}). Since the upper layer of this heterojunction is a homogeneous grating, it can slide adiabatically at a velocity $\nu$ without breaking the invariance of the system under the transformation $\delta\rightarrow\delta+\Lambda$.

The spectrum of this photonic heterojunction is obtained using the Finite-Difference Time-Domain (FDTD) method implemented in the solver 3D Electromagnetic Simulator of the commercial software Lumerical -- see Fig.~\ref{fig:edge}({b}). As the upper layer slides to the right, a distinct mode, absent in the spectra of either config.~(1) or~(2) individually, traverses the spectral gap from one band to the other. In the vicinity of $t=0.5T$, it sharply contrasts with other modes for having a consistent descent in wavelength. Conversely, if the upper layer slides to the left, this mode's wavelength monotonically increases. This phenomenon is known for the soliton mode in the Rice-Mele model~\cite{Asboth2016,Citro2023} and can be interpreted as a chiral edge mode along the synthetic dimension~\cite{Minh2023}. Indeed, by plotting the field distribution of this mode in Fig.~\ref{fig:edge}({d}), we see that it strongly localizes at the interface of the heterojunction and exponentially decays into the constituent crystals with mismatched decay lengths. The decay lengths differ because the dispersions of the two crystals are different. This interface mode is robust against any perturbations that preserve the bulk spectral gap of both sides, e.g., see Fig.~S10 of the SI.

Since the interface mode acts as a cavity that confines electromagnetic field, we quantify this confinement by the mode's quality (Q) factor. Figure~\ref{fig:edge}({e}) shows the Q-factor of the interface mode at various moments for two heterojunctions of different sizes. For the smaller heterojunction, this quantity varies greatly over time, or equivalently, the interlayer displacement~$\delta$: it becomes smallest when the interface mode lies at the center of the spectral gap. This dependence wanes as the system size increases, and the Q-factor can attain values as high as $10^5$. This suggests potential applications of this photonic heterojunction in devices, such as lasers, beam emitter~\cite{Lee2022beam}, and filters. A demonstration of this photonic heterojunction as a filter is provided in the SI.

\subsection{Dynamical Control of Topological Phases}
While the wavelength of the topological interface mode is continuously tunable via the adjustment of the lateral displacement $\delta$, this mode can also be switched on and off by inducing a topological transition on one side of the heterojunction. Using the picture of two competing potentials, we see that in the left side of the heterojunction~[Fig.~\ref{fig:edge}(a)] the optical mode is pulled by a moving potential of the upper layer. If the potential of the lower layer is sufficiently enhanced to dominate over the upper one, the left side transitions into the trapping regime while the right side's regime remains unchanged. With both sides of the heterojunction being in the same regime, the interface mode vanishes.

Motivated by a recent observation of topological phase transition in photonic crystal using a PCM~\cite{Uemura2024}, we demonstrate this idea in our bilayer lattice by incorporating a PCM into its design~\cite{Cao2019,Tian2023,Han2024,Quan2024}. Specifically, the lower grating can be fabricated using amorphous antimony trisulfide ($\text{Sb}_2\text{S}_3$)~\cite{Faneca2020}, an earth-abundant and nontoxic PCM with ultralow losses~\cite{Delaney2020,Gutierrez2022}. $\text{Sb}_2\text{S}_3$ is stable at room temperature in both its amorphous ($n_2=2.73$) and crystalline ($n_2=3.26$) phases, which can be changed reversibly by either heating the entire sample or selectively illuminating it with laser pulses. The change in refractive index of the lower layer when $\text{Sb}_2\text{S}_3$ transitions alters the coupling strengths between the guided waves~\cite{Bentata2025}, enhancing the stationary potential created by the lower layer. Thus, it may induce a topological phase transition in the bilayer photonic crystal, switching it between the pumping and trapping regimes (see Materials and Methods for a complete topological phase diagram). This phenomenon indeed occurs in our current photonic heterojunction where, upon the crystallization of $\text{Sb}_2\text{S}_3$, the photonic crystal on the left side of the heterojunction changes from the pumping to trapping regime while the right side remains in the trapping regime. The spectrum of the lowest spectral gap in the vicinity of $t=0.5T$ is shown in Fig.~\ref{fig:edge}({c}) when $\text{Sb}_2\text{S}_3$ crystallizes completely. The gap widens and shifts to longer wavelength compared to Fig.~\ref{fig:edge}({b}) since the refractive index of the lower grating increases. Importantly, the interface mode traversing the gap disappears as the two sides of the heterojunction now share the same topology, which serves as a clear signature of the topological phase transition. This also illustrates how PCMs can be used to dynamically switch on and off an optical interface mode.
\subsection{Topological Lasing}
\begin{figure*}
	\includegraphics[width=\linewidth]{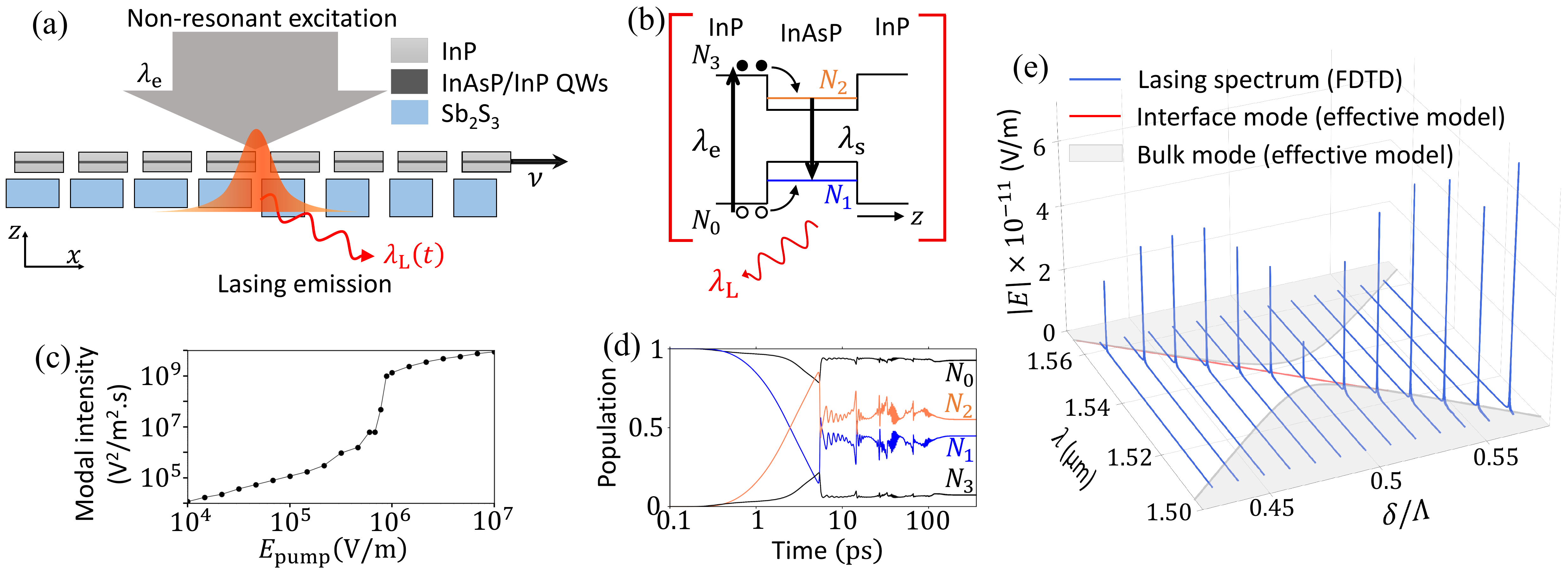}
	\caption{\label{fig:lasing} \textbf{Reconfigurable topological lasing.} (a) The heterojunction of bilayer photonic crystal where the upper layer is made of gain material InAsP/InP moves slowly with velocity $\nu$. The heterojunction is continuously pumped by a non-resonant source of wavelength $\lambda_e=850$~nm and achieves lasing action at $\lambda_L$. (b) Schematic diagram of the four-level model for the lasing action with the spontaneous emission wavelength $\lambda_s=1500$~nm. (c) Dependence of modal intensity of the lasing signal at $\delta=0.5\Lambda$ on the pump electric field strength. (d) Evolution in time of electron populations at the four levels at $E_{\text{pump}}=\SI{1.3e6}{\volt\per\meter}$. (e) Spectrum of the lasing mode at several moments $t$ when the upper layer slides and the non-resonant source has the same strength as (c). The gray line and areas indicate the interface and bulk modes following the effective model. The FDTD simulations are performed with $N=150$.}
\end{figure*}
Owing to the high Q-factor and strong confinement of our photonic heterojunction, we can incorporate active (i.e., gain) materials into one layer and perform non-resonant optical excitation to induce lasing at the interface mode~\cite{Ota2018,Han2019,Gong2020}. The lasing wavelength can be selectively and continuously varied through the dynamical sliding motion of the upper layer. In particular, we use the design of a photonic junction presented before [Fig.~\ref{fig:edge}({a})] with the upper grating composed of InP and InAsP quantum wells (QWs). This layer is continuously illuminated by a non-resonant optical source of wavelength $\lambda_\text{e}=\SI{850}{\nano\meter}$ [Fig.~\ref{fig:lasing}({a})]. As sketched in Fig.~\ref{fig:lasing}({b}), the pump injects hot carriers to the conduction and valence bands of InP. These hot electrons and holes then relax to the fundamental states of the QWs and then recombine radiatively, leading to a spontaneous emission centered at wavelength $\lambda_\text{s}=\SI{1500}{\nano\meter}$. In the spontaneous emission, only photons of wavelength $\lambda_{\text{L}}$ associated with the interface mode are in resonance and confined within the cavity; those of other wavelengths decay rapidly. After the electronic population inversion is established, lasing emission can be achieved precisely at the wavelength $\lambda_\text{L}$.

We numerically validate this idea through FDTD simulations with the gain material modeled by a four-level two-electron material~\cite{Chang2004}. Two levels with electron populations $N_0$ and $N_3$ are the band edges of the barrier InP while $N_1$ and $N_2$ are two levels of the quantum well InAsP [Fig.~\ref{fig:lasing}({b})]. We examine the lasing action with the two layers displaced by $\delta=0.5\Lambda$. By increasing the field strength of the pump, we observe an emission peak at $\lambda_{\text{L}}$ for $E_{\text{pump}}\gtrsim7\times10^5$~V/m. The dependence of the lasing modal intensity on the pump field strength is shown in Fig.~\ref{fig:lasing}({c}) with the characteristic laser threshold behavior -- a clear transition from spontaneous to stimulated emission. Here, the modal intensity is defined by $\frac{1}{2\pi}\int d\omega |E|^2$ with the integration taken over the frequency range $\omega$ encompassing the lasing peak. The temporal evolution of the electron populations at field strength $\SI{1.3e6}{\volt\per\meter}$ is shown in Fig.~\ref{fig:lasing}({d}), where the population inversion between level 1 and 2 is achieved. The steady state is reached at around 0.16~ns after the excitation and the gain material becomes transparent, giving rise to lasing at the edge mode.

Finally, translating the upper layer slowly at a fixed pump intensity yields the emission spectrum at various moments $t$ shown in Fig.~\ref{fig:lasing}({e}), where the lasing peaks align with the interface mode. The variation of the lasing wavelength with respect to time is locked to the sliding direction of the upper layer. The spectrum obtained from the effective model is presented as a visual reference showing the interface mode's variation; it matches perfectly with the FDTD simulation of Fig.~\ref{fig:edge}({b}). A slight discontinuity at $\delta=0.5\Lambda$, accompanied by a dip in the emission field strength, is present due to the finite size of the simulated structure (see the SI for more remarks). The presence of spontaneous emission in the spectrum at $t\lesssim0.5T$ means that the lasing threshold varies with respect to the relative displacement $\delta$. Single-mode lasing across a range of wavelengths can be achieved by constantly keeping the pump power above the threshold. This demonstrates the tunability of the lasing mode in this photonic heterojunction. Such a lasing mode is robust against defects and disorders since they are topologically protected. The single-mode operation is guaranteed as the number of edge mode is one, which is dictated by the change of Chern number across the heterojunction.
\section{Conclusion}
Regarding the experimental feasibility, the bilayer photonic crystal can be fabricated using standard nanofabrication methods, such as electron beam lithography and ionic dry etching~\cite{saadi2025,zhuang2025,jing2025,Tang2023}. Dynamic control over the vertical and lateral degrees of freedom can be facilitated by MEMSs~\cite{Wu2006,Liang2014vi,Chollet2016,Ren2019}. Especially, a recent MEMS integrated into a bilayer photonic lattice has demonstrated its capability to dynamically tune various degrees of freedom, including the interlayer spacing, relative rotation, lateral translation, tilting, and stretching~\cite{Tang2025}. The operation speed of MEMS is negligible compared with the speed of light, guaranteeing the adiabatic pumping of the lasing mode. The phase of the PCM $\text{Sb}_2\text{S}_3$ can be reversibly switched using state of the art microheaters~\cite{Wen2024}, such as indium-tin-oxide heater~\cite{Hemmatyar2021}, silicon PIN diode heater~\cite{Chen2023}, or graphene-based heater~\cite{Fang2022}.

Our proposal of combining MEMS and PCMs for dynamically controlling topological interface modes demonstrates the potential of this bilayer photonic heterojunction for realizing multidimensionally reconfigurable photonic devices. As applications, this includes lasers, beam emitters and filters, providing unprecedented mechanisms of creating and manipulating light. Fundamentally, our results also lay the groundwork for further investigations into 2D Thouless pumping, which is connected to the 4D quantum Hall effect~\cite{Kraus2013}, and for examining Thouless pumping in moir\'{e} lattices, as predicted in twisted bilayer graphene~\cite{Fujimoto2020, Su2020} and noted by Thouless himself~\cite{Thouless1983}. Furthermore, this study opens avenues for exploring novel aspects of Thouless pumping beyond the adiabatic regime~\cite{Liu2025shortcut} and even in the relativistic regime, where the grating motion approaches the speed of light~\cite{Horsley2023, Horsley2024}.
\section{Experimental/Methods}
\subsection*{PWE simulations}
The PWE simulations in this work are carried out by the MIT Photonic Bands package~\cite{MPB} with a 2D computational cell of size $(L_x,L_z) = (1,5)\Lambda$. The resolution is 64 pixels per $\Lambda$. The number of bands computed is eight.
\subsection*{FDTD simulations}
The FDTD simulations in this work are carried out by either the MEEP package~\cite{MEEP} or the commercial software Lumerical~\cite{Lumerical}.

{\it \textbf{Spectrum} --} The spectra shown in Figs.~\ref{fig:edge}(b) and~\ref{fig:edge}(c) of the photonic heterojunction are obtained from Lumerical FDTD simulations. A dielectric photonic junction is constructed following the geometry depicted in Fig.~\ref{fig:edge}(a) with its interface lying at the center of the computational cell. The refractive indices of the upper and lower gratings are 3.17 and 2.73, respectively. In this linear regime, the parameters scale with the lattice constant $\Lambda$, so we set $\Lambda=\SI{1}{\micro\meter}$ for simplicity. The total number of periods is 400, i.e., the length of the heterojunction is $\SI{400}{\micro\meter}$. The 2D computational cell is enclosed in standard phase-matching layers. The mesh for finite-difference calculation has the maximum mesh step $\SI{0.02}{\micro\meter}$ along the $x$ direction and 67 mesh cells per wavelength along the $z$ direction. The electromagnetic modes of the system are excited by 20 electric dipoles randomly distributed in the bilayer within a range of $\SI{160}{\micro\meter}$ around the interface. The dipoles are aligned along the $y$ axis ($\theta=0$), have random phases and random angle with respect to the $x$ axis. Each of them emits a broadband pulse with frequency ranging from 69~THz to 74~THz. The simulation runs for $\SI{70}{\pico\second}$ at 300~K. All signals are recorded and analyzed by 20 time monitors randomly distributed in the system within a range of $\SI{240}{\micro\meter}$ around the interface. We note that the spectra in Fig.~\ref{fig:edge} have a few discrete patterns. They are numerical artifacts stemming from the dielectric gratings crossing a mesh line.\\

{\it \textbf{Quality factor} --} The quality factor of the edge mode is computed using MEEP and Lumerical simulations, with both methods yielding comparable results. In the MEEP simulations, a dielectric photonic junction is constructed similar to that in Lumerical. A single point source, emitting a Gaussian pulse with a frequency width of $\Delta f = 0.002(c/\Lambda)$, is randomly embedded in a dielectric rod at the interface. The central frequency of this optical pulse follows a straight trajectory along the chiral edge mode, $f_{\text{center}}=(0.06\delta + 0.2086)(c/\Lambda)$. The source excites modes with an electric field parallel to the dielectric rod. A monitor is placed inside another dielectric rod at the junction interface to analyze the response for $10^4$ time units after the source has turned off. The 2D computational cell has a resolution of 32 and dimensions of $(N+7,26)$, where $N$ is the number of periods on each side. The boundary layers perpendicular to the $y$-axis are phase matching layers of thickness 2, while those normal to the $x$-axis are adiabatic absorbers of thickness 7. The periodic lattices submerge into the absorbers.\\

{\it \textbf{Lasing simulation} --} For lasing simulations in Lumerical FDTD, the heterojunction is constructed similarly but we use a realistic geometry with $\Lambda=\SI{366}{\nano\meter}$ since the calculations are nonlinear. The lower grating is a dielectric with the refractive index $2.73$ while the upper grating now is modeled by a 4-level 2-electron material~\cite{Chang2004}, akin to what is depicted in Fig.~\ref{fig:lasing}(b). In this gain material, the transition wavelengths are $\lambda_{\text{s}} = \SI{1.5}{\micro\meter}$ and $\lambda_{\text{e}} = \SI{0.85}{\micro\meter}$, the damping coefficients are $\gamma_{\text{a}}=\gamma_{\text{b}}=10^{13}$~Hz, the lifetimes of different decay channels are $t_{30}=t_{21}=3\times10^{-10}$~s and $t_{32}=t_{10}=10^{-13}$~s, and the electron population density is $\SI{1e23}{\per\meter\cubed}$. The heterojunction is continuously pumped by a spatial-Gaussian beam with wavelength $\lambda_{\text{e}}$ and waist radius $\SI{2}{\micro\meter}$, located $\SI{2.2}{\micro\meter}$ above the system. The signals are recorded and analyzed by 10 time monitors located $\SI{0.5}{\micro\meter}$ below the system. The heterojunction in these simulations has 300 periods, corresponding to a length of approximately $\SI{110}{\micro\meter}$. The simulations run for 360~ps at 300~K. The mesh of the finite-difference method has the maximum mesh step $\SI{0.006}{\micro\meter}$ along the $x$ direction and 60 mesh cells per wavelength along the $z$ direction.
More details about the lasing simulations can be found in the SI.
\section*{Acknowledgments}
We acknowledge Pierre Viktorovitch for having participated in the early stage of the project. We are grateful to Pierre Delplace, David Carpentier, S\'{e}bastien Cueff, and Alexander Cerjan for fruitful discussions. This work was granted access to the HPC resources of PMCS2I (P\^{o}le de Mod\'{e}lisation et de Calcul en Sciences de l’Ing\'{e}nieur de l’Information) of \'{E}cole Centrale de Lyon, \'{E}cully, France. The authors acknowledge the technical and human support provided by the DIPC Supercomputing Center. We acknowledge the assistance of AI language models Gemini and ChatGPT in correcting the grammatical errors of this manuscript, and Scholar Labs in extending the literature review.

\bibliography{bilayer_bib}

\section*{Funding Sources}
This work was supported by\\
Spanish Ministerio de Ciencia, Innovaci\'{o}n y Universidades grant PRE2021-097126 (DHMN)\\
Deutsche Forschungsgemeinschaft (DFG, German Research Foundation, project numbers 447948357 and 440958198), Sino-German Center for Research Promotion (Project M-0294), the ERC (Consolidator Grant 683107/TempoQ), German Ministry of Education and Research (Project QuKuK, BMBF Grant No. 16KIS1618K), and the EIN~Quantum~NRW (HCN)\\
Brown Theoretical Physics Center, Institute for Basic Science in Korea through the Project IBS-R024-D1 (DXN)\\
French National Research Agency grant ANR-17-CE24-0020, IDEXLYON from Université de Lyon, Scientific Breakthrough project TORE within the Programme Investissements d'Avenir grant ANR-19-IDEX-0005 (TL, XL, HSN) \\
French National Research Agency (ANR) under the project POLAROID (ANR-24-CE24-7616-01) (HSN) \\
Transnational Common Laboratory \emph{QuantumChemPhys}, Department of Education of the Basque Government through the project PIBA 2023 1 0007 (STRAINER), Basque Foundation for Science (IKERBASQUE), PID2020-120614GB-I00 (ENACT) funded by the Spanish MICIU/AEI and by ERDF/EU, "Artificial Quantum Matter: From 2D Materials to Spin Lattice Systems – Programa Fundamentos de la Fundación BBVA 2024" (DB)

\section*{Author Contributions}
Conceptualization: HSN, DXN, HCN, DHMN, DB \\
Investigation: DHMN, HCN, TL\\
Methodology: XV, HCN, DXN\\
Software: ED, XV, TL\\
Supervision: HSN, DB\\
Writing -- original draft: HCN \\
Writing -- review \& editing:  DHMN, DB, HSN, DXN

\let\addcontentsline\oldaddcontentsline

\onecolumngrid
\pagebreak

\renewcommand{\theequation}{S\arabic{equation}}
\renewcommand{\thefigure}{S\arabic{figure}}
\setcounter{equation}{0}
\setcounter{section}{0}
\setcounter{figure}{0}
\setcounter{page}{1}

\begin{center}
	\large{\textbf{SUPPLEMENTAL MATERIAL:}}\\
	\Large{\textbf{Topological Lasing from Thouless Pumping in Bilayer Photonic Crystal}}

\bigbreak
	
\small D.-H.-Minh~Nguyen$^{\ast}$, Dung~Xuan~Nguyen$^{\ast}$, Hai-Chau~Nguyen, Thibaud~Louvet,
Emmanuel~Drouard, Xavier~Letartre, Dario~Bercioux$^{\ast}$,
Hai~Son~Nguyen$^{\ast}$\\
\end{center}

\tableofcontents

\onecolumngrid

\section{Effective Hamiltonian}
In this section, we present the derivation of the effective Hamiltonian. The system consists of two gratings with the same period $\Lambda$ separated by distance $D$. The confinement of electromagnetic waves within each grating is considerably analogous to the problem of an electron in a finite quantum well. Hence, for simplicity, we will investigate how the optical modes in each grating are effectively described and then phenomenologically add the evanescent coupling between modes in different gratings.

We consider a symmetric grating of dielectric constant $\varepsilon_s$ in the air with thickness $H$ and width $L$, the spatial dielectric function is given by $\varepsilon(x,z) = [\varepsilon(x)-1]f_{\text{e}}(z) + 1$, where the relative dielectric constant of the grating with respect to the environment is a periodic function with the period $\Lambda$ and 
\begin{equation}
	\varepsilon(x) = \left[\begin{array}{rcl}
		\varepsilon_s & \text{for } -L/2<x<L/2 \\
		1 & \text{for } -\Lambda/2<x<-L/2\text{ or } L/2<x<\Lambda/2
	\end{array}\right. ,
\end{equation}
and $f_{\text{e}}(z) = \Theta(z+W/2)-\Theta(z-W/2)$~\footnote{$\Theta(x)$ is the Heaviside function.}. As $\varepsilon(x)-1$ is a periodic function, we can write its Fourier expansion as $\varepsilon(x)-1 = \sum_{n=-\infty}^{+\infty}\xi_ne^{i2\pi nx/\Lambda}$, which gives
\begin{equation}
	\varepsilon(x,z) = \xi_0f_{\text{e}}(z) + 1 + \sum_{n\neq0}\xi_n(z)e^{i\frac{2\pi n}{\Lambda}x} = \bar{\varepsilon}(z) + \sum_{n\neq0}\xi_n(z)e^{i\frac{2\pi n}{\Lambda}x},\qquad \xi_n(z) = \xi_nf_{\text{e}}(z).
\end{equation}
The electromagnetic field of this system is governed by the Maxwell's equations
\begin{align*}
	\nabla\cdot\mathbf{H} = 0,&\qquad \nabla\times\mathbf{H} = \varepsilon_0\varepsilon\frac{\partial\mathbf{E}}{\partial t},\nonumber\\
	\nabla\cdot(\varepsilon\mathbf{E}) = 0,&\qquad \nabla\times\mathbf{E} = -\mu_0\frac{\partial\mathbf{H}}{\partial t}.
\end{align*}
Since the system is uniform and infinite along the $y$ direction, we can decompose the solutions at momentum $k_y=0$ into two sets of modes: transverse electric (TE) modes with $E_x=E_z=0$, and transverse magnetic (TM) modes with $H_x=H_z=0$. These modes have fields' strengths distribute uniformly along the $y$ direction. In our case, we are only interested in TE modes but a theory for the TM ones can be developed similarly. The Maxwell's equations are consequently reduced to the wave equation
\begin{equation}
	\frac{\partial^2 E_y}{\partial z^2} + \frac{\partial^2 E_y}{\partial x^2} = -\varepsilon(x,z)\frac{\omega^2}{c^2} E_y. 
\end{equation}
Due to the discrete translation symmetry of the grating, we employ the Bloch theorem to write the electric field in terms of plane waves
\begin{equation}
	E_y(\mathbf{r}) = \sum_nC_n(z)e^{i(k_x+K_n)x},\qquad\qquad \text{with } K_n=\frac{2\pi n}{\Lambda}.
\end{equation}
Inserting this expression into the wave equation, we get
\begin{align}
	\sum_ne^{i(k_x+K_n)x}\left[\frac{\partial^2}{\partial z^2} + \bar{\varepsilon}(z)\frac{\omega^2}{c^2} - (k_x+K_n)^2\right]C_n(z) = -\frac{\omega^2}{c^2}\sum_{n,l\neq0}\xi_l(z)e^{i(k_x+K_{n+l})x}C_n(z).
\end{align}
Multiply two sides with $e^{-i(k_x+K_m)x}$ and integrate over $x$, we get
\begin{align}
	\left[\frac{\partial^2}{\partial z^2} + \bar{\varepsilon}(z)\frac{\omega^2}{c^2} - (k_x+K_n)^2\right]C_n(z) = -\frac{\omega^2}{c^2}\sum_{l\neq n}\xi_{n-l}(z)C_l(z).
\end{align}
We divide the plane wave basis into two sets: basic waves $\mathcal{B}$ and others $\mathcal{O}$. The basic waves are those that contribute most to the modes of interest. We impose our first approximation by assuming that the basic waves have the form $C_n(z) = E_{0y}(z)C_n$ for $n\in\mathcal{B}$, where $E_{0y}(z)$ is the envelop function of the lowest-frequency guided TE modes in a homogeneous slab described by the dielectric function $\bar{\varepsilon}(z)$. We have
\begin{equation}
	\frac{\partial^2 E_{0y}(z)}{\partial z^2} - (k_x+K_n)^2E_{0y}(z) = -\bar{\varepsilon}(z)\frac{\omega^2}{c^2}E_{0y}(z).
\end{equation}
Here, $\omega$ and $E_{0y}(z)$ vary with respect to the momentum $k_x+K_n$. Since in the case of grating, we only work with a small range of momentum around a high-symmetry point $k_x+K_n=\tilde{K}$, we assume that $\omega$ and $E_{0y}(z)$ are $k-$independent and are values at this point $\tilde{K}$. Thus, these two quantities are now determined via the wave equation
\begin{equation}
	\frac{\partial^2 E_{0y}(z)}{\partial z^2} - \tilde{K}^2E_{0y}(z) = -\bar{\varepsilon}(z)\frac{\omega^2_{\tilde{K}}}{c^2}E_{0y}(z).
\end{equation}
Combining this equation with the wave equation for electromagnetic waves in the grating, we arrive at
\begin{align}
	\left[\bar{\varepsilon}(z)\frac{\omega^2-\omega_{\tilde{K}}^2}{c^2} + \tilde{K}^2 - (k_x+K_n)^2\right]E_{0y}(z)C_n = -\frac{\omega^2}{c^2}\left[E_{0y}(z)\sum_{l\in\mathcal{B},l\neq n}\xi_{n-l}(z)C_l + \sum_{u\in\mathcal{O}}\xi_{n-u}(z)C_u(z)\right].
\end{align}
Then, multiplying both sides by $E_{0y}^*(z)$ and taking integration over $z$ yields
\begin{align}
	\left\{\omega^2-\omega_{\tilde{K}}^2 + \frac{c^2}{\bar{n}_0^2}\left[\tilde{K}^2 - (k_x+K_n)^2\right]\right\}C_n = -\alpha\omega^2\sum_{l\in\mathcal{B},l\neq n}\xi_{n-l}C_l - \omega^2\sum_{u\in\mathcal{O}}\xi_{n-u}\int_{-\infty}^{+\infty}dzE_{0y}^*(z)C_u(z)f_{\text{e}}(z)
\end{align}
with $\bar{n}_0^2 = \displaystyle\int_{-\infty}^{+\infty}\left|E_{0y}(z)\right|^2\bar{\varepsilon}(z)dz$ and $\alpha = \displaystyle\int_{-\infty}^{+\infty}\left|E_{0y}(z)\right|^2f_{\text{e}}(z)dz$. We choose the point of interest $\tilde{K}=K_1/2$, notate $\omega_{\tilde{K}}=\omega_0$, and consider two basic waves corresponding to $n=0$ and $n=-1$, which gives two coupled equations
\begin{subequations}
	\begin{align}
		\left\{\omega^2-\omega_0^2 + \frac{c^2}{\bar{n}_0^2}\left[\frac{K_1^2}{4} - (k_x+K_{-1})^2\right]\right\}C_{-1} &= -\alpha\omega^2\xi_{-1}C_0 - \omega^2\sum_{u\neq0,-1}\xi_{-1-u}\int_{-\infty}^{+\infty}dzE_{0y}^*(z)C_u(z)f_{\text{e}}(z),\\
		\left\{\omega^2-\omega_0^2 + \frac{c^2}{\bar{n}_0^2}\left(\frac{K_1^2}{4} - k_x^2\right)\right\}C_0 &= -\alpha\omega^2\xi_1C_{-1} - \omega^2\sum_{u\neq0,-1}\xi_{-u}\int_{-\infty}^{+\infty}dzE_{0y}^*(z)C_u(z)f_{\text{e}}(z).
	\end{align}
\end{subequations}
Define $k = k_x - K_1/2$ and neglect the interaction with higher-order modes ($u\neq0,-1$), we obtain
\begin{subequations}
	\begin{align}
		\left\{\omega^2-\omega_0^2 + \frac{c^2}{\bar{n}_0^2}\left[\frac{K_1^2}{4} - \left(k-\frac{K_1}{2}\right)^2\right]\right\}C_{-1} &= -\alpha\omega^2\xi_{-1}C_0\Rightarrow \left[\omega^2-\omega_0^2 - \frac{c^2}{\bar{n}_0^2}\left(k^2 - K_1k\right)\right]C_{-1} = -\alpha\omega^2\xi_{-1}C_0,\\
		\left\{\omega^2-\omega_0^2 + \frac{c^2}{\bar{n}_0^2}\left[\frac{K_1^2}{4} - \left(k+\frac{K_1}{2}\right)^2\right]\right\}C_0 &= -\alpha\omega^2\xi_1C_{-1}\Rightarrow \left[\omega^2-\omega_0^2 - \frac{c^2}{\bar{n}_0^2}\left(k^2 + K_1k\right)\right]C_0 = -\alpha\omega^2\xi_1C_{-1}.
	\end{align}
\end{subequations}
If the periodic modulation in the dielectric function of the grating is sufficiently weak, we can considerably simplify these equations by assuming that $\boxed{|\omega-\omega_0|\ll\omega_0}$ and neglecting the term of $(\omega-\omega_0)\xi_{\pm1}$. The equations then become
\begin{subequations}
	\begin{align}
		\left[\omega-\omega_0 - \frac{c^2}{2\bar{n}_0^2\omega_0}\left(k^2 - K_1k\right)\right]C_{-1} \approx -\frac{\alpha\omega_0\xi_{-1}}{2}C_0,\\
		\left[\omega-\omega_0 - \frac{c^2}{2\bar{n}_0^2\omega_0}\left(k^2 + K_1k\right)\right]C_0 \approx -\frac{\alpha\omega_0\xi_1}{2}C_{-1}.
	\end{align}
\end{subequations}
Rewriting these two equations in matrix form gives us the effective model of a single grating
\begin{equation}
	\omega_0 + \frac{c^2}{2\bar{n}_0^2\omega_0}k^2 +
	\begin{pmatrix}
		\bar{v}k & U \\ U^\ast & -\bar{v}k
	\end{pmatrix}
	\begin{pmatrix}
		C_0 \\ C_{-1}
	\end{pmatrix}
	= \omega \begin{pmatrix} C_0 \\ C_{-1} \end{pmatrix}
\end{equation}
with $\bar{v} = \dfrac{\pi c^2}{\bar{n}_0^2\omega_0\Lambda}$ and $U = -\dfrac{\alpha\omega_0\xi_1}{2}$. Owing to the inversion symmetry along the $x$ direction of the dielectric function, we have $U=U^*$. For simplicity, we neglect the $k$-quadratic term as it has no effect on the topological properties of the spectrum -- the wave equation can thus be written as an eigen-equation $H_{\text{mono}}(k)\Psi(k) = \omega_k\Psi(k)$ with the operator
\begin{equation}
	H(k) = \omega_0 +
	\begin{pmatrix}
		\bar{v}k & U \\ U & -\bar{v}k
	\end{pmatrix}\label{Eq: single grating Hamiltonian}
\end{equation}
termed the Hamiltonian and $\Psi(k) = \begin{pmatrix} C_0 & C_{-1} \end{pmatrix}^T$.

\begin{figure}
	\includegraphics[width=\linewidth]{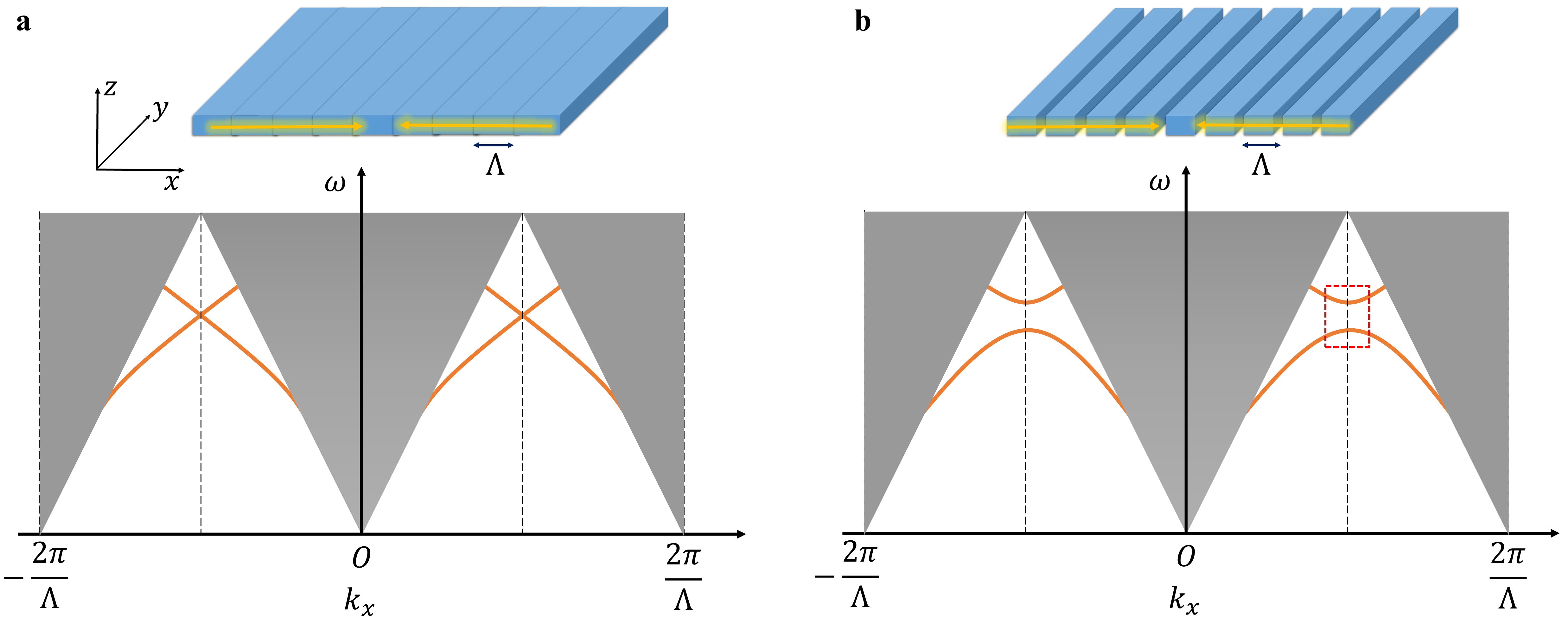}
	\caption{\label{fig: explain model} Dispersion of the lowest guided mode in (a) a homogeneous waveguide with infinitesimal periodic modulation of dielectric, and (b) a dielectric grating. The red dashed box indicates the region where the effective model is valid.}
\end{figure}

The basis functions of this Hamiltonian consists of two plane waves $\varphi_+ = E_{0y}(z)e^{ik_xx} = E_{0y}(z)e^{i(k + \pi/\Lambda)x}$ and $\varphi_- = E_{0y}(z)e^{i(k_x - 2\pi/\Lambda)x} = E_{0y}(z)e^{i(k - \pi/\Lambda)x}$. The electric field is hence given by
\begin{equation}
	E_y(\mathbf{r}) = E_{0y}(z)\left[C_0e^{i(k + \pi/\Lambda)x} + C_{-1}e^{i(k - \pi/\Lambda)x}\right].
\end{equation}
This effective model can be interpreted as depicted in Fig.~\ref{fig: explain model}. In a homogeneous slab waveguide with infinitesimal periodic modulation of dielectric constant, two counter-propagating guided modes have group velocity $\bar{v}$ when their wave numbers are around $k_x=\pi/\Lambda$ ($X$ point). In the presence of periodic corrugation, these modes diffract and couple with each other with strength $U$.
If the grating translates along the $x$ axis by $\delta$, its dielectric function is given by $\varepsilon'(x) = \varepsilon(x-\delta)$. While the zeroth Fourier component $\xi_0'$ remains unchanged, the first component varies as follows
\begin{align}
	\xi_1' = \int_{-\infty}^{+\infty}dx\left[\varepsilon'(x) - 1\right]e^{-i2\pi x/\Lambda} = \int_{-\infty}^{+\infty}dx\left[\varepsilon(x-\delta) - 1\right]&e^{-i2\pi x/\Lambda} \nonumber\\
	&= \int_{-\infty}^{+\infty}dx\left[\varepsilon(x) - 1\right]e^{-i2\pi(x+\delta)/\Lambda} = \xi_1e^{-i2\pi\delta/\Lambda}.
\end{align}
As a result, we arrive at the substitution $U\rightarrow Ue^{-i2\pi\delta/\Lambda}$.

For the bilayer grating, we can follow a similar procedure to obtain the effective Hamiltonian. However, for simplicity, we treat the problem phenomenologically by assuming that the co-propagating modes in the two layers couple with each other only through the evanescent field. Other interlayer coupling mechanisms are negligible. With the evanescent coupling strength notated $V$, we achieve the effective Hamiltonian shown in the main text
\begin{equation}
	\label{eq:Hamiltonian}
	H(k,\delta) = \begin{pmatrix} 
		\omega_1+v_1k & U_1e^{ -i 2\pi \frac{\delta}{\Lambda}} & V &0\\
		U_1e^{ i 2\pi \frac{\delta}{\Lambda}} & \omega_1-v_1k &0 &V\\
		V &0 & \omega_2+v_2k & U_2 \\
		0& V & U_2 & \omega_2-v_2k	
	\end{pmatrix},
\end{equation}
whose eigenvalues $\omega(k,q)$ are the frequencies of the four lowest guided modes in the vicinity of the $X$-point
\begin{equation}
	H(k,\delta)\begin{pmatrix} C_0^u \\ C_{-1}^u \\ C_0^l \\ C_{-1}^l \end{pmatrix} = \omega(k,q)\begin{pmatrix} C_0^u \\ C_{-1}^u \\ C_0^l \\ C_{-1}^l \end{pmatrix}.
\end{equation}
Here, the indices $u$ and $l$ denote the upper grating and lower grating, respectively. The electric field of these TE modes is given by
\begin{equation}
	E_y(\mathbf{r}) = E_{0y}^u(z)\left[C_0^ue^{i(k + \pi/\Lambda)x} + C_{-1}^ue^{i(k - \pi/\Lambda)x}\right] + E_{0y}^l(z)\left[C_0^le^{i(k + \pi/\Lambda)x} + C_{-1}^le^{i(k - \pi/\Lambda)x}\right].\label{eq: original basis waves}
\end{equation}
For short notations, we write $E_y(\mathbf{r}) = \sum_{\nu=1}^4\Psi_\nu\varphi_\nu$ for $\Psi_\nu\in\{C_0^u, C_{-1}^u, C_0^l, C_{-1}^l\}$ and $$\varphi_\nu\in\left\{E_{0y}^u(z)e^{i(k + \pi/\Lambda)x}, E_{0y}^u(z)e^{i(k - \pi/\Lambda)x}, E_{0y}^l(z)e^{i(k + \pi/\Lambda)x}, E_{0y}^l(z)e^{i(k - \pi/\Lambda)x}\right\}.$$

\begin{figure}
	\begin{center}
		\includegraphics[width=1 \textwidth]{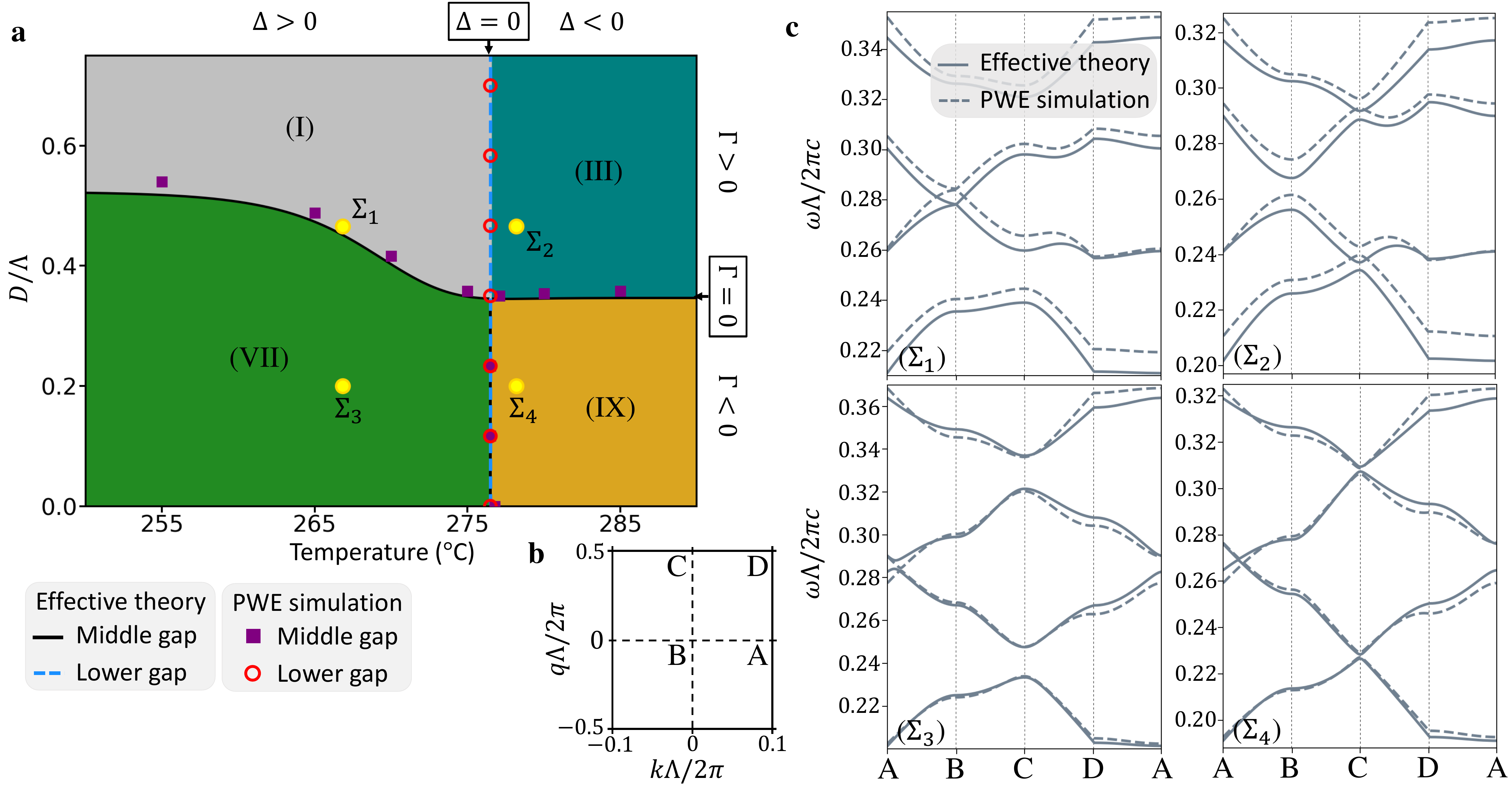}
		\caption{Comparison between the bulk band structure obtained from PWE simulations and the one calculated by the effective theory. (a) The phase diagram presented in the main text. (b) The contour ABCDA in momentum space that is used to plot the band structures. (c) The four band structures at the four $\Sigma$-points in (a).}
		\label{fig:Supp_phase}
	\end{center}
\end{figure}

The parameters of the Hamiltonian can be retrieved through comparison with rigorous simulations at some special points, as exemplified in Fig.~\ref{fig:param_retrieved} for silicon gratings. With these parameters, the four frequency bands can be straightforwardly obtained by exact diagonalization and are shown in Fig.~\ref{fig:Full_Dispersion}. The effective model indeed agrees excellently with Finite-Difference Time-Domain (FDTD) simulations. We carry out the same fitting process for our setup in the main text.
From Figs.~\ref{fig:Supp_phase}(c), we see that the effective model again agrees with PWE simulations. We note that in case $(\Sigma_3)$ of Fig.~\ref{fig:Supp_phase}(c), the two middle bands obtained from PWE simulation has a gap at the AB segment which is small compared to the one given by the effective theory.
\section{Dynamic Control of Topological Phases via Phase-Change Material}
\begin{figure}
	\centering
	\includegraphics[width=0.5\linewidth]{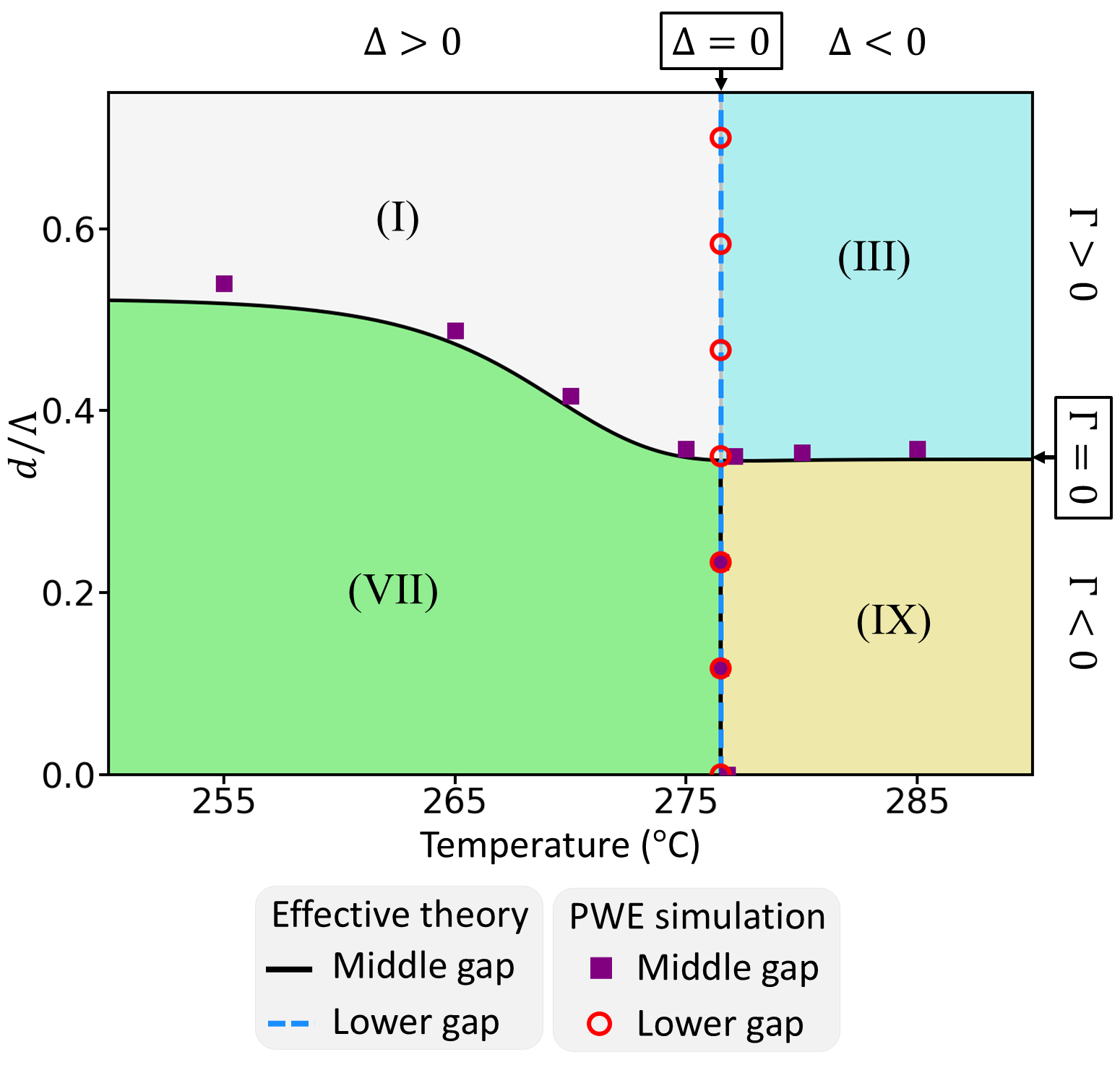}
	\caption{\label{fig:phase} \textbf{Phase diagram controlled by varying the temperature and interlayer distance.} The bilayer comprises an antimony trisulfide grating and an amorphous silicon grating. The solid black (dashed blue) line denotes where the middle (lower) gap closes within the effective theory. The purple squares (red circles) correspond to similar points obtained through PWE simulations.}
\end{figure}

The topological phases presented previously can be observed in a single sample by means of PCMs~\cite{Cao2019,Han2024,Uemura2024,Quan2024}. In this work, we incorporate into the bilayer photonic crystal the PCM antimony trisulfide ($\text{Sb}_2\text{S}_3$), which is well known for its ultralow losses across the visible and near-infrared wavelengths in both crystalline and amorphous phases~\cite{Delaney2020,Gutierrez2022}, and its exceptional tunability~\cite{Hemmatyar2021,Laprais2024}. Its phase can be changed reversibly, either by heating the entire sample or by shining laser pulses at a specific spot.

In our bilayer system, as the intralayer coupling strength depends on the refractive index, we vary $\Delta$ by switching the material's phase between amorphous and crystalline. We demonstrate this idea by designing a bilayer system of an amorphous silicon (refractive index 3.15) grating and an $\text{Sb}_2\text{S}_3$ grating. The geometrical parameters are $w_1=w_2=0.8\Lambda$ and $h_1=h_2=0.37\Lambda$. The refractive index of $\text{Sb}_2\text{S}_3$ increases continuously from 2.73 to 3.26 when it transitions from amorphous to crystalline phase. This transition is complete when the temperature is raised above $280^{\circ}\text{C}$~\cite{Taute2023simple}. Amorphization can be achieved by heating the crystalline $\text{Sb}_2\text{S}_3$ above its melting temperature and then rapidly quenching. An on-chip reversible transition of PCM can be obtained using a state of the art microheater~\cite{Hemmatyar2021,Abdollahramezani2022,Fang2022,Chen2023,Wen2024}. Here, by fitting the experimental data describing the temperature-dependent refractive index of $\text{Sb}_2\text{S}_3$ during its crystallization~\cite{Taute2023simple}, we obtained a function analogous to the logistic function
\begin{equation*}
	n_{\text{Sb}_2\text{S}_3} = A + B\left\{1 - \frac{1}{1 + 0.5\left[e^{\alpha(T-T_0)}+e^{\beta(T-T_0)}\right]}\right\}
\end{equation*}
with $A=2.732738$, $B=0.530212$, $\alpha=\SI{0.47062}{\per\kelvin}$, $\beta=\SI{0.23360}{\per\kelvin}$, and $T_0=\SI{273}{\kelvin}$. The PWE simulations are run accordingly using this function. The dependence of parameters $\omega_2$, $v_2$, and $U_2$ of the PCM grating and the interlayer coupling $V$ on temperature is achieved by fitting the effective model of the monolayer and bilayer lattices with PWE results. The experimental data of $\text{Sb}_2\text{S}_3$ and the temperature dependence of the effective model's parameters are presented in the SM.

On the other hand, to alter $\Gamma$, the interlayer distance can be adjusted dynamically using on-chip MEMSs~\cite{Tang2023,Tang2024hBN}. Such a combination of thermal and mechanical control of the bilayer grating allows us to achieve the complete phase diagram shown in Fig.~\ref{fig:phase} on a single sample, corresponding to $\text{Sb}_2\text{S}_3$ crystallization. The four regions are associated with the four gapped states, i.e., the four frequency bands are disconnected, while their borders correspond to the gapless ones. These gap-closing lines match well with the results obtained from the PWE simulation using MIT Photonic Bands package~\cite{MPB}.

The phase diagram is general and can be achieved with other dielectric materials as long as the geometrical parameters are appropriate. As considered above, despite replacing amorphous silicon with indium phosphide (InP), whose refractive index is 3.17, all the topological phases remain. This generality implies possible optimization of the bilayer grating for specific properties, such as the spectral gap or the quality factor of the heterostructure.
\section{Temperature Dependence of Antimony Trisulfide}
\begin{figure}
	\includegraphics[width=\linewidth]{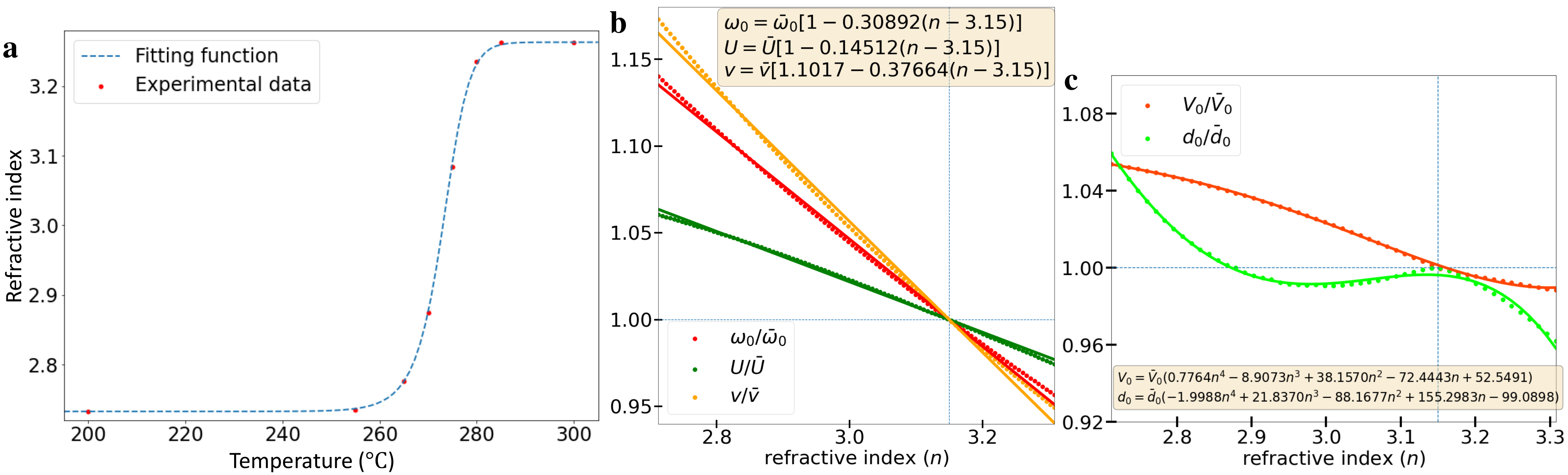}
	\caption{\label{fig:pcm} (a) The dependence of $\text{Sb}_2\text{S}_3$ refractive index on temperature. (b) Variation of model parameters of a single grating ($L=0.8\Lambda$ and $H=0.37\Lambda$) with respect to the refractive index. (c) Variation of interlayer coupling strength and decay length with respect to the refractive index.}
\end{figure}

Antimony trisulfide ($\text{Sb}_2\text{S}_3$) is an ultra-low loss phase-change material. The dependence of its refractive index on temperature for progressive crystallization is presented in the table below, which was provided by the authors of Ref.~\cite{Taute2023simple}
\begin{center}
	\begin{tabular}{ |c|c|c|c|c|c|c|c|c| } 
		\hline
		\multicolumn{9}{|c|}{Refractive index for wavelength 1500 nm}\\
		\hline
		Temperature ($\SI{}{\degreeCelsius}$) & $200$ & $255$ & $265$ & $270$ & $275$ & $280$ & $285$ & $300$ \\ 
		\hline
		Refractive index & $2.732738$ & $2.735401$ & $2.776021$ & $2.87417$ & $3.084485$ & $3.235228$ & $3.26294$ & $3.26295$ \\ 
		\hline
	\end{tabular}
\end{center}
The continuous change in refractive index arises from gradual partial crystallization. The apparition of nucleation sites of $\text{Sb}_2\text{S}_3$ is reported to have perfectly random and homogeneous distribution~\cite{Taute2023simple}. We fit these data with a function analogous to the logistic function
\begin{equation}
	n_{\text{Sb}_2\text{S}_3} = A + B\left\{1 - \frac{1}{1 + 0.5\left[e^{\alpha(T-T_0)}+e^{\beta(T-T_0)}\right]}\right\}
\end{equation}
with $A=2.732738$, $B=0.530212$, $\alpha=\SI{0.47062}{\per\kelvin}$, $\beta=\SI{0.23360}{\per\kelvin}$, and $T_0=\SI{273}{\kelvin}$, as shown in Fig.~\ref{fig:pcm}(a).

To obtain the phase diagram in Fig.~7, we retrieve the dependence of model parameters on the refractive index of the $\text{Sb}_2\text{S}_3$ grating. First, we consider a single $\text{Sb}_2\text{S}_3$ grating with $L=0.8\Lambda$ and $H=0.37\Lambda$ and see how the model parameters vary. The simulation data are obtained by PWE with MPB package. We crudely fit the data with linear functions in comparison with parameters ($\bar{\omega}_0, \bar{U}$, and $\bar{v}$) of an identical grating made of amorphous silicon (aSi), as shown in Fig.~\ref{fig:pcm}(b). Then, we examine a bilayer of $\text{Sb}_2\text{S}_3$ grating and aSi grating with the same geometrical parameters. The interlayer interaction between them is approximated as $V(D) = V_0\exp(-D/d_0)$. The dependence of $V_0$ and $d_0$ is fit with polynomial functions -- see Fig.~\ref{fig:pcm}(c). Here, $\bar{V}_0$ and $\bar{d}_0$ correspond to the case when $\text{Sb}_2\text{S}_3$ refractive index is 3.15, the same as amorphous silicon.

The parameters are
\begin{equation*}
	\frac{\bar{\omega}_0\Lambda}{2\pi c^2} = 0.26304,\qquad \frac{\bar{U}\Lambda}{2\pi c^2} = 0.02181,\qquad \bar{v} = 0.36222c,\qquad \frac{\bar{V}_0\Lambda}{2\pi c^2} = 0.06067,\qquad \bar{d}_0 = 0.35599\Lambda.
\end{equation*}
\section{Guided Transmission for Broadband Filtering}
In this section, we demonstrate that a heterojunction of the bilayer grating can be used as a tunable filter.
\subsection*{Design}
\begin{figure}
	\begin{center}
		\includegraphics[width=\textwidth]{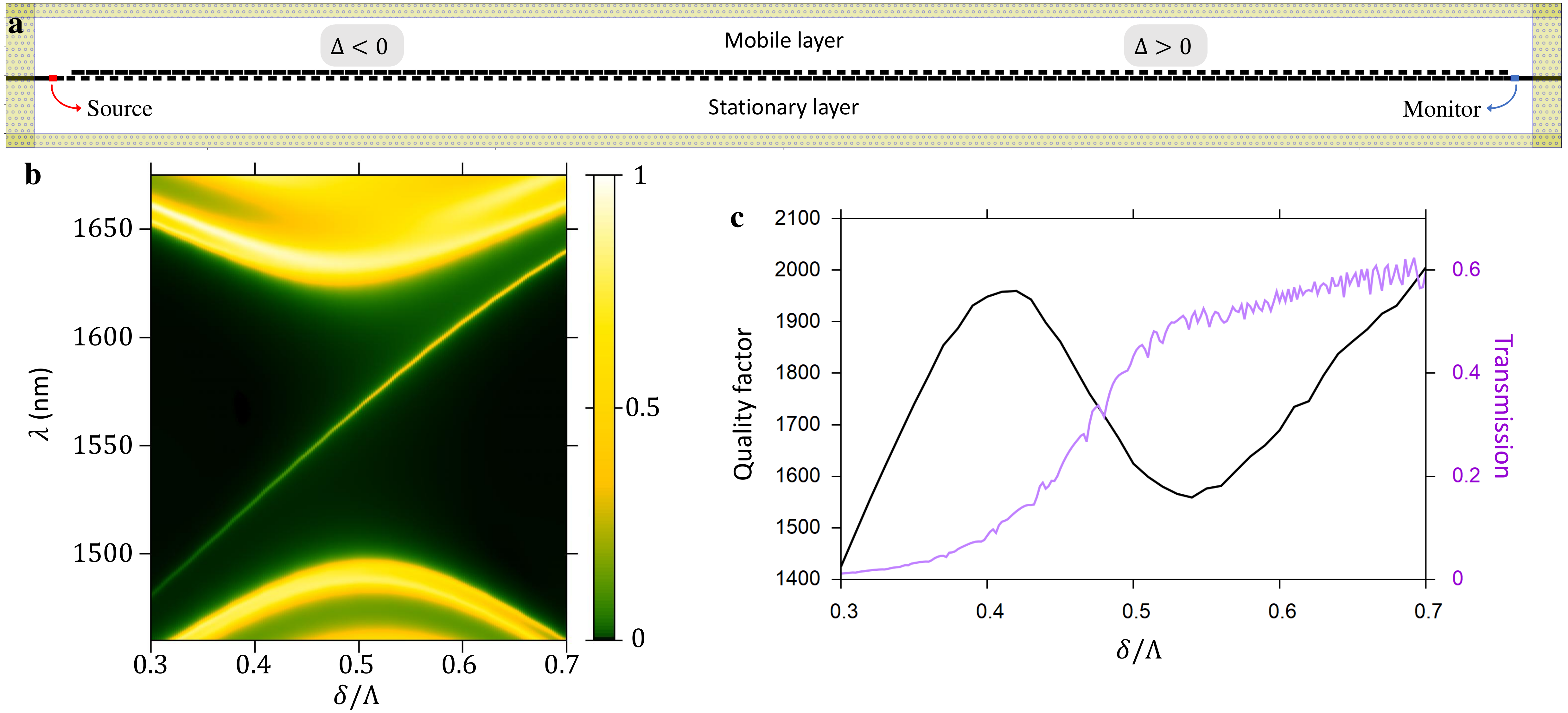}
		\caption{Photonic junction as a light filter. (a) FDTD computational cell of the heterojunction, which consists of two bilayer gratings with opposite values of $\Delta$. The cell is surrounded by phase-matching layers that dissipates the electromagnetic field. (b) The guided transmission spectrum of the system as the upper grating translates gradually. (c) The quality factor and transmission coefficient of the interface mode.}
		\label{fig:filter}
	\end{center}
\end{figure}

The design of the light filter is shown in Fig.~\ref{fig:filter}(a), which is a photonic heterojunction composed of two aligned bilayer gratings on two sides. Each bilayer system consists of two silicon gratings (refractive index 3.5) with identical thickness $H=0.3\Lambda$ and different values of width: $L_1=0.9\Lambda$ and $L_2=0.58\Lambda$. The interlayer separation between the two gratings is $0.1\Lambda$. The two sides of the junction share the same structure but the two gratings are swapped, which reverses the sign of $\Delta$ across the junction. The lower layer is attached stock-still to a source and a monitor, both of which are embedded in a silicon padding block, while the upper layer is mobile and used as a ``tunable knob".

The number of lattice periods per side of the junction is 20. The source emits a pulse with Gaussian shape in frequency, centering at $0.218c/\Lambda$ and of width $0.03c/\Lambda$. The FDTD simulations are carried out using MEEP.
\subsection*{Guided transmission}
From the main text, we know that a localized interface mode exists in this junction due to the topological phase transition, and it is chiral along the synthetic dimension $\delta$. Importantly, as seen in Fig.~4D, this mode exponentially decays into the bulk -- we can thus excite this mode by putting a source in its decaying tail. The upper layer is translated along the $x$ direction to tune the frequency of the edge mode by varying $\delta$.

The transmission spectrum for various values of $\delta$ is shown in Fig.~\ref{fig:filter}(b) with lattice period $\Lambda=340$~nm. We see that the system in this case can filter a wide range of wavelength, from approximately 1450~nm to 1650~nm\footnote{As the spectral gap has not been optimized in this example, the range of filtering wavelength can certainly be further increased.}. The quality factor of the transmitted signal is shown in Fig.~\ref{fig:filter}(c), which is above 1000, implying the excellent performance of light filtering even for a wide band gap. This quality factor becomes greater when the system size gets larger, i.e., larger $N$, as we have seen in Fig.~4E. However, as the source is further away from the interface, the transmitted signal decreases exponentially. We can optimize the system size to obtain the desirable output as decreasing the size decreases the quality factor but increases the transmission intensity. Additionally, Fig.~\ref{fig:filter}(c) also presents the transmission of the edge mode, which depends strongly on the shift $\delta$.
\subsection*{Advantages}
Besides being a high-quality filter over a wide range of wavelength, this photonic junction is also a filter robust against disorders and defects. Since the chiral edge mode is topologically protected in the synthetic momentum space, it consistently traverses the spectral gap even in the presence of perturbations~\cite{Minh2023}. Consequently, with the relative displacement $\delta$ being dynamically adjustable, one can always tune the edge mode to achieve the desirable wavelength.
\section{Finite-Difference Time-Domain Simulations}
All the FDTD simulations in this work are carried out by either the MEEP package~\cite{MEEP} or the commercial software Lumerical.

{\it \textbf{Spectrum} --} The spectra shown in Figs.~4(b) and 4(c) of the photonic heterojunction are obtained from Lumerical FDTD simulations. A dielectric photonic junction is constructed following the geometry depicted in Fig.~4(a) with its interface lying at the center of the computational cell. The refractive indices of the upper and lower gratings are 3.17 and 2.73, respectively. In this linear regime, the parameters scale with the lattice constant $\Lambda$, so we set $\Lambda=\SI{1}{\micro\meter}$ for simplicity. The total number of periods is 400, i.e., the length of the heterojunction is $\SI{400}{\micro\meter}$. The 2D computational cell is enclosed in standard phase-matching layers. The mesh for finite-difference calculation has the maximum mesh step $\SI{0.02}{\micro\meter}$ along the $x$ direction and 67 mesh cells per wavelength along the $y$ direction (i.e., $z$ direction in Results). The electromagnetic modes of the system are excited by 20 electric dipoles randomly distributed in the bilayer within a range of $\SI{160}{\micro\meter}$ around the interface. The dipoles are aligned along the $z$ axis ($\theta=0$), have random phases and random angle with respect to the $x$ axis. Each of them emits a broadband pulse with frequency ranging from 69~THz to 74~THz. The simulation runs for $\SI{70}{\pico\second}$ at 300~K. All signals are recorded and analyzed by 20 time monitors randomly distributed in the system within a range of $\SI{240}{\micro\meter}$ around the interface. We note that the spectra in Fig.~4 have a few discrete patterns. They are numerical artifacts resulting from the dielectric gratings crossing a mesh line.\\

\begin{figure}[h!]
	\begin{center}
		\includegraphics[width=1 \textwidth]{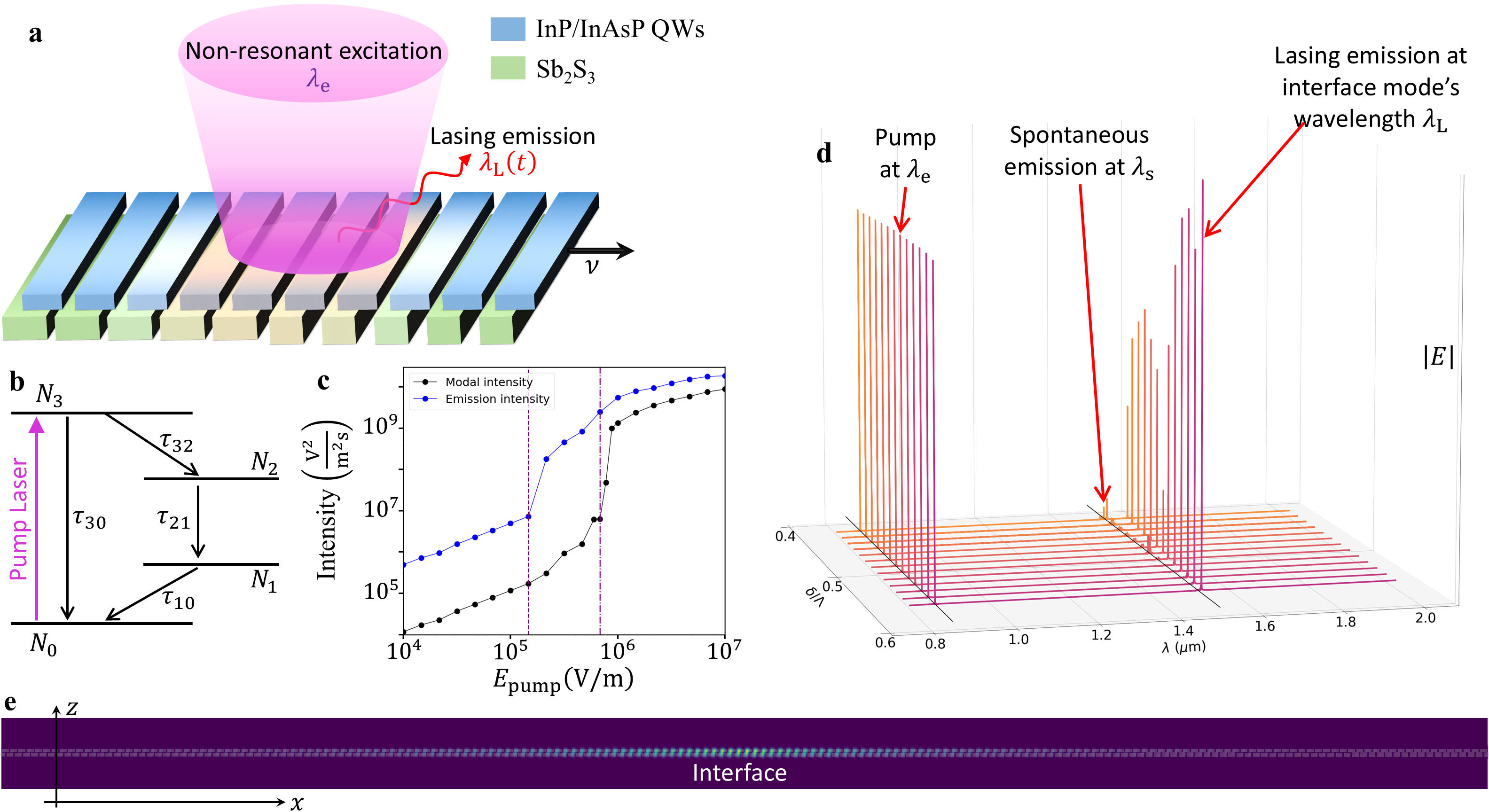}
		\caption{(a) The heterojunction of bilayer photonic crystal where the upper layer is made of gain material InAsP/InP moves slowly with velocity $\nu$. The heterojunction is continuously pumped by a non-resonant spatially Gaussian source of wavelength $\lambda_\text{e}$ and achieves lasing action at $\lambda_\text{L}$. (b) Schematic diagram of the four-level two-electron model describing the gain material. (c) Emission and modal intensities with respect to the pump field strength when $\delta=0.5\Lambda$. The dashed line indicates where population inversion starts to take place while the dashed dotted line indicates the lasing threshold. (d) Complete spectrum detected at the monitors and (e) the electric field profile of the lasing mode at $\delta=0.5\Lambda$ when $E_{\text{pump}} = \SI{1.3e6}{\volt\per\meter}$.}
		\label{fig:lasing model}
	\end{center}
\end{figure}

{\it \textbf{Quality factor} --} The quality factor of the edge mode is computed using MEEP and Lumerical simulations, with both methods yielding comparable results. In the MEEP simulations, a dielectric photonic junction is constructed similar to that in Lumerical. A single point source, emitting a Gaussian pulse with a frequency width of $\Delta f = 0.002(c/\Lambda)$, is randomly embedded in a dielectric rod at the interface. The central frequency of this optical pulse follows a straight trajectory along the chiral edge mode, $f_{\text{center}}=(0.06\delta + 0.2086)(c/\Lambda)$. The source excites modes with an electric field parallel to the dielectric rod. A monitor is placed inside another dielectric rod at the junction interface to analyze the response for $10^4$ time units after the source has turned off. The 2D computational cell has a resolution of 32 and dimensions of $(N+7,26)$, where $N$ is the number of periods on each side. The boundary layers perpendicular to the $y$-axis are phase matching layers of thickness 2, while those normal to the $x$-axis are adiabatic absorbers of thickness 7. The periodic lattices submerge into the absorbers.\\

{\it \textbf{Lasing simulation} --} For lasing simulations in Lumerical FDTD, the heterojunction is constructed similarly but we use a realistic geometry with $\Lambda=\SI{366}{\nano\meter}$ since the calculations are nonlinear. The lower grating is still a dielectric with the refractive index $2.73$ while the upper grating now is modeled by a four-level two-electron material~\cite{Chang2004}, akin to what is depicted in Fig.~5(b). In this gain material, the transition wavelengths are $\lambda_{\text{s}} = \SI{1.5}{\micro\meter}$ and $\lambda_{\text{e}} = \SI{0.85}{\micro\meter}$, the damping coefficients are $\gamma_{\text{a}}=\gamma_{\text{b}}=10^{13}$~Hz, the lifetimes of different decay channels are $t_{30}=t_{21}=3\times10^{-10}$~s and $t_{32}=t_{10}=10^{-13}$~s, and the electron population density is $\SI{1e23}{\per\meter\cubed}$. The heterojunction is continuously pumped by a spatial-Gaussian beam with wavelength $\lambda_{\text{e}}$ and waist radius $\SI{2}{\micro\meter}$, located $\SI{2.2}{\micro\meter}$ above the system. The signals are recorded and analyzed by 10 time monitors located $\SI{0.5}{\micro\meter}$ below the system. The heterojunction in these simulations has 300 periods, corresponding to a length of approximately $\SI{110}{\micro\meter}$. The simulations run for 360~ps at 300~K. The mesh for finite-difference calculation has the maximum mesh step $\SI{0.006}{\micro\meter}$ along the $x$ direction and 60 mesh cells per wavelength along the $y$ direction.
\begin{figure}
	\begin{center}
		\includegraphics[width=1 \textwidth]{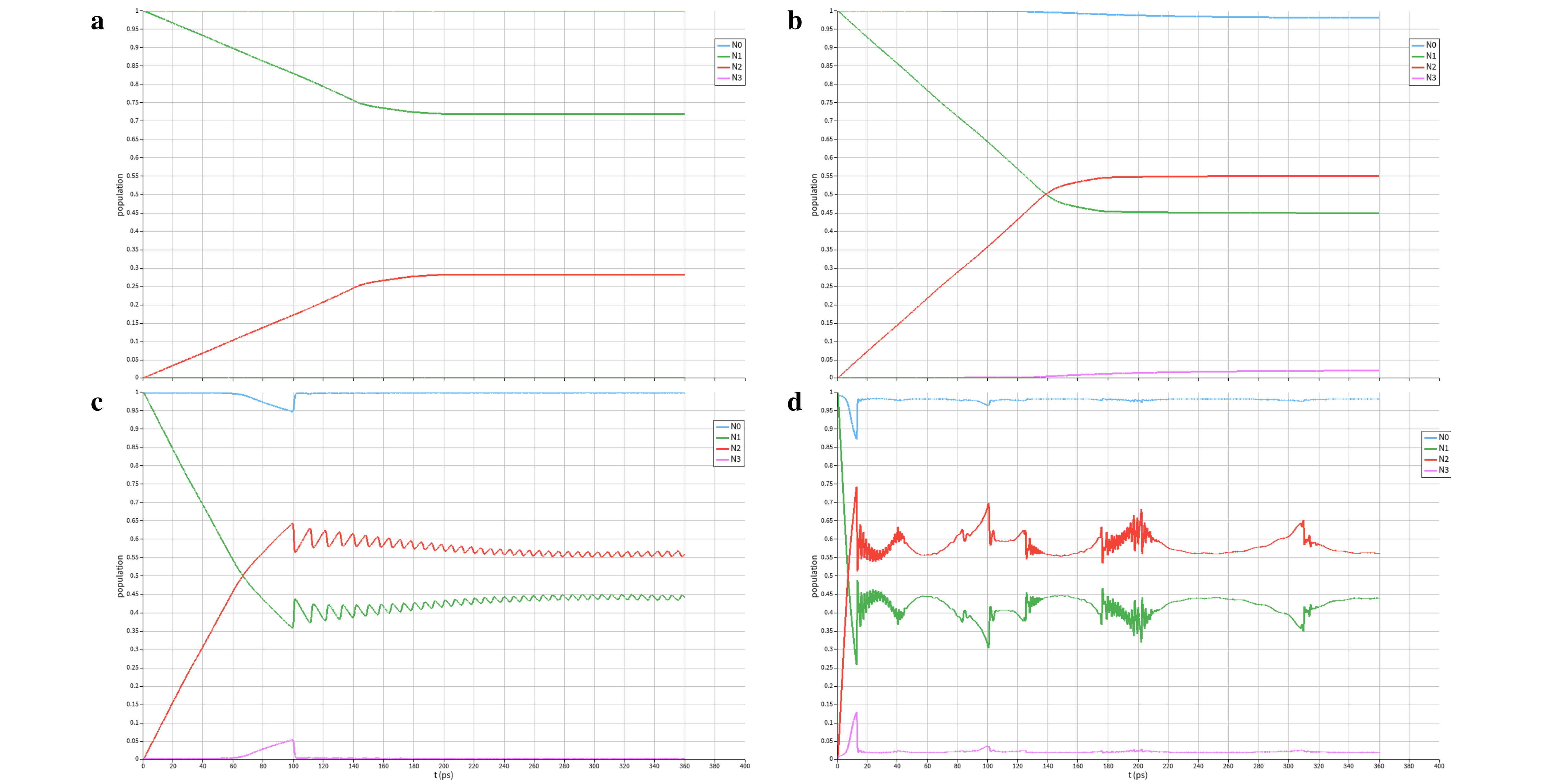}
		\caption{Electron population density probability of the four levels at different pump field strengths: (a) $E_{\text{pump}} = 10^5$~V/m, (b) $E_{\text{pump}} = 146780$~V/m, (c) $E_{\text{pump}} = 215440$~V/m, (d) $E_{\text{pump}} = 316230$~V/m.}
		\label{fig:lasing population}
	\end{center}
\end{figure}

In our lasing simulation with Lumerical FDTD - the setup is re-sketched in Fig.~\ref{fig:lasing model}(a), the gain material is described by a 4-level 2-electron model, which is depicted in Fig.~\ref{fig:lasing model}(b). In this model, the electron transitions are treated as two coupled dipole oscillators, one corresponds to levels 1 and 2 while the other is associated with levels 0 and 3. These transitions are governed by the coupled rate equations and the Pauli exclusion principle, and they are solved self-consistently. At $t=0$, the electron populations are $N_0=N_1=1$ and $N_2=N_3=0$. Other parameters are given in the Materials and Methods.

We remark on Fig.~5(e) of the main text, where we observe a slight discontinuity in the emission signals along the chiral interface mode and a corresponding dip in intensity. This stems from the coupling between the central interface mode and an unphysical interface mode at the boundary of the computational cell, caused by the finite-size effects of the structure and the finite thickness of the absorbing layers. Such in-plane leakage of the interface mode leads to the surge at $\delta=0.5\Lambda$ in the Q-factor shown in Fig.~4(e). On the other hand, we note that the signal detected by the monitors depends not only on the Q-factor of the cavity but also on the positions of the source and the monitors. In our current simulation, the monitors are located directly below the interface, which do not capture the entire lasing emission. For instance, if the angular dependence of the lasing emission varies with respect to the lateral shift $\delta$, the signal's intensity will vary accordingly, independent of the Q-factor. Hence, the variation of the emission intensity with respect to $\delta$ detected at the monitors is a combined function of the Q-factor and the spatial configuration of the source and monitors.

When examining the lasing action with respect to the field strength of the pump at $\delta=0.5\Lambda$, there are two ways to present the emission signal. The first one is to compute the lasing modal intensity, which is shown in the main text. Here, we focus solely on the intensity of light at the lasing wavelength. The second way is to compute the emission intensity, which is also defined by $\frac{1}{2\pi}\int d\omega |E|^2$ but with the integration taken over the frequency range $\omega$ encompassing all the emission peaks. Both of these quantities are shown in Fig.~\ref{fig:lasing model}(c) against the pump field strength. On the one hand, we see that the emission intensity depicts nicely where the population inversion (between level 1 and 2) starts to take place, which is $E_{\text{inv}}\approx1.5\times10^5$~V/m. Between $E_{\text{inv}}$ and $E_{\text{thres}}$ is where amplified spontaneous emission dominates, i.e., population inversion is present with no distinguishable signal at the resonant wavelength. This process does not involve any resonator, e.g., a cavity, and has a broad bandwidth in its emission spectrum, centered at $\lambda_{\text{s}}=1500$~nm. It is expected to take place here owing to the large volume of gain material. The electron populations at some values of field strengths around this transition are shown in Fig.~\ref{fig:lasing population}. On the other hand, the modal intensity illustrates well the lasing threshold $E_{\text{thres}}\approx7\times10^5$~V/m where the lasing peak starts to appear. As shown in the main text and Fig.~\ref{fig:lasing model}(d) for field strength $E_{\text{pump}}=\SI{1.3e6}{\volt\per\meter}$ above the threshold, when population inversion as well as steady state are achieved, the lasing mode appears as a sharp peak at the wavelength $\lambda_{\text{L}}$ of the localized interface mode. It varies linearly against $\delta$ due to the chiral nature, different from the signals from the pumping source and spontaneous emission, whose wavelengths always center at $\lambda_\text{e}$ and $\lambda_\text{s}$, respectively -- see Fig.~\ref{fig:lasing model}(d). We further confirm that this signal indeed comes from the topological interface mode by plotting its field profile in Fig.~\ref{fig:lasing model}(e), which localizes at the interface, in agreement with the interface mode's profile.
\section{Edge State from Effective Model}
In Figure~5D of the main text, we use the effective model to compute the chiral edge mode within the synthetic space, which serves as a guide for the simulation results. The method for this calculation is detailed in the Supplemental Material of Ref.~\cite{Minh2023}.
\section{Captions of Supplemental Videos}
\subsection*{Caption for Movie S1}
\textbf{Particle pumping and trapping in bipartite potential}. Time evolution of two periodic potentials of the same periodicity where $U_1(x,t)$ moves slowly and $U_2(x) = 1.5\sin(2\pi x/\Lambda)$ is stationary, with $U_1(x,t) = 2.2\sin(2\pi x/\Lambda - 2\pi\nu t/\Lambda)$ (left) and $U_1(x,t) = 1.2\sin(2\pi x/\Lambda - 2\pi\nu t/\Lambda)$ (right). The particles (denoted by black dots) can either be transported by $U_1(x,t)$ to the next unit cell (pumping), or be pulled back by $U_2(x)$ to their original positions (trapping).
\subsection*{Caption for Movie S2}
\textbf{Optical pumping and trapping in bilayer photonic lattice}. Time evolution of the electric field profile of the lowest mode at $k=0$ at various moments in the sliding bilayer photonic lattice. The two configurations of the lattice are associated with two regimes: topological pumping and trapping. The electric field is computed using the plane-wave expansion method implemented in the MIT photonic bands package.
\section{Supplemental Figures}
\begin{figure}[h!]
	\begin{center}
		\includegraphics[width=0.8 \textwidth]{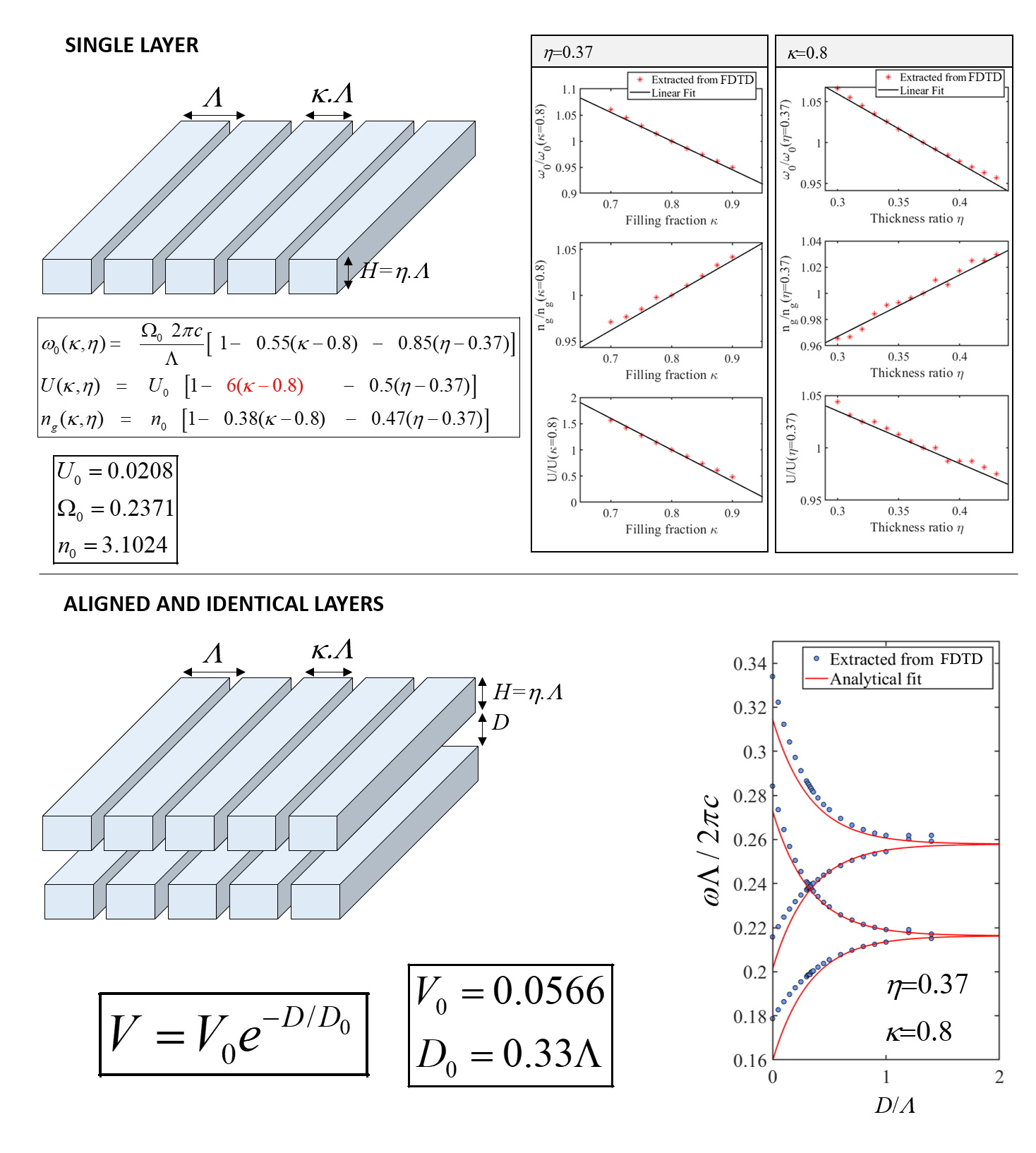}	
		\caption{Parameter retrievals for the effective Hamiltonian from FDTD simulations. Upper Panels: the simulations for a single layer grating are used to obtain the dependence of $\omega_0$, $U$ and $n_g$ as functions of the filling fraction $\kappa$ and thickness ratio $\eta$. These parameters are extracted by fitting the two band dispersion with the Hamiltonian~\eqref{Eq: single grating Hamiltonian}. Lower Panels: the simulations for a bilayer of aligned and identical gratings at $k=0$ are used to obtain the dependence of $V$ as functions of the distance $D$ separating the two layers. The exponential decay law of $V$ is obtained by fitting the four band edges $\omega_0 + U + V$, $\omega_0 + U - V$, $\omega_0 - U + V$, and $\omega_0 - U - V$ at varying distance $D$. Here, the values of $U$ and $\omega_0$ are already known from the retrievals from the single layer simulations.}
		\label{fig:param_retrieved}
	\end{center}
\end{figure}

\begin{figure}
	\begin{center}
		\includegraphics[width=0.9\textwidth]{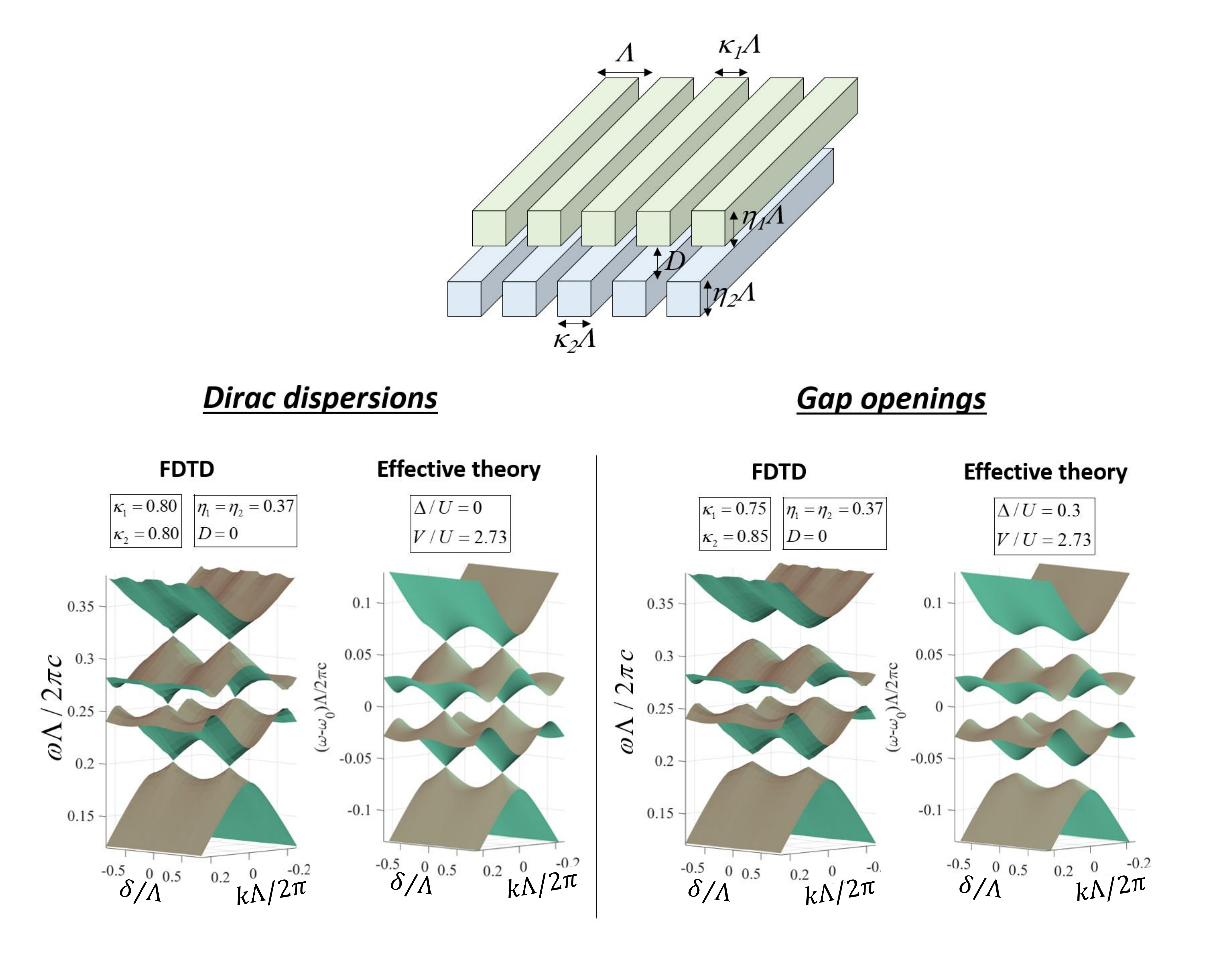}
		\caption{Comparison between the bulk band structure obtained from FDTD simulations and the one calculated by the effective theory using the retrieved parameters shown in Fig.~\ref{fig:param_retrieved}. Here, the simulated band structures are only shown in the vicinity of $k=0$ (i.e., $X$ point of the 1D Brillouin zone) since the effective theory is only valid in this region.}
		\label{fig:Full_Dispersion}
	\end{center}
\end{figure}

\begin{figure}
	\begin{center}
		\includegraphics[width=1 \textwidth]{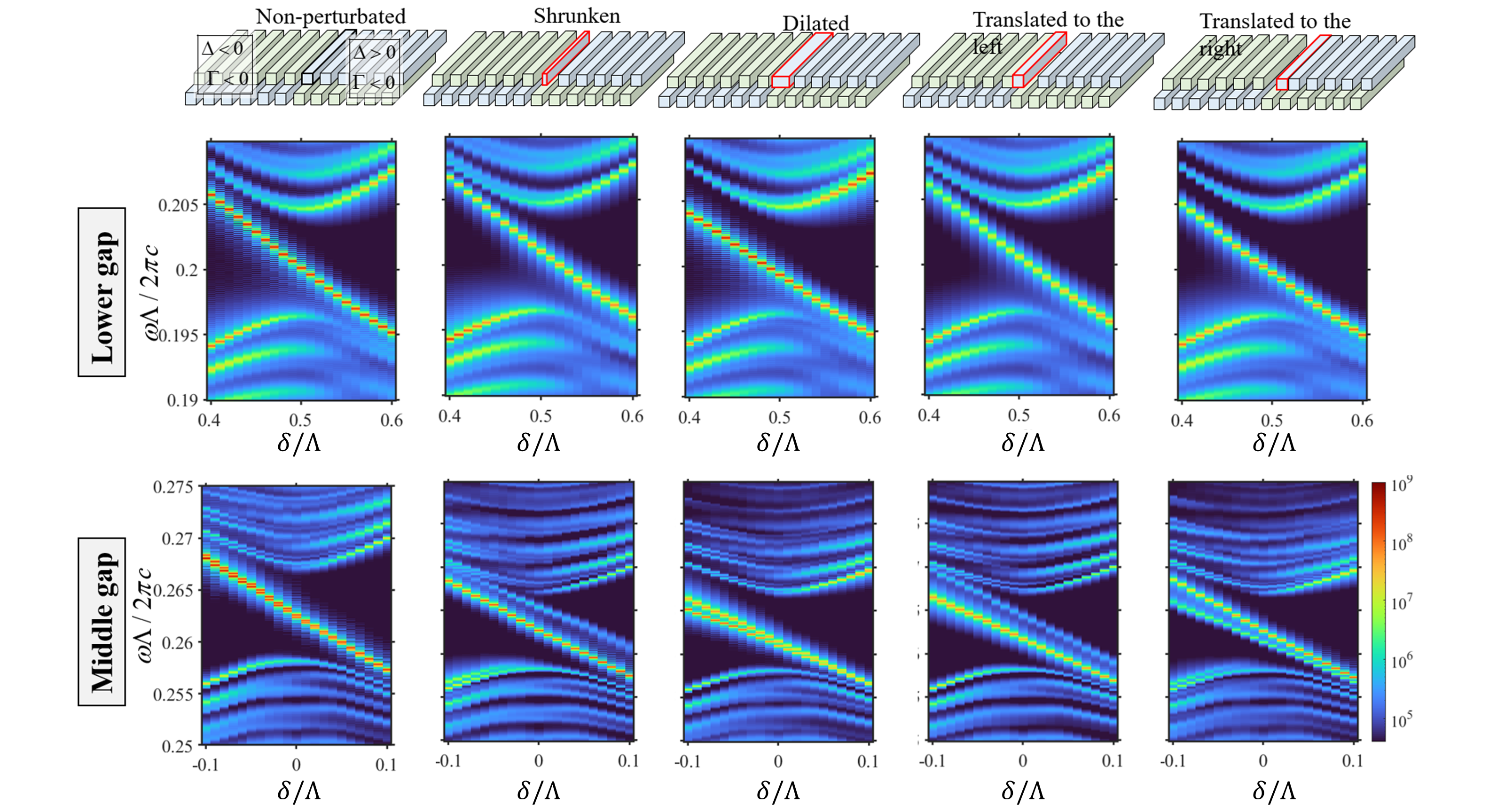}	
		\caption{The robustness of the chiral edge states against defects at the interface. Here, the interface is strongly perturbed by modifying either the size or the position of the first rod in the upper-right grating. These results show that: i) the chirality is topologically protected against perturbations, and ii) the degeneracy lifting of the two edge states in the middle gap depends strongly on the interface, i.e., the boundary conditions.}
		\label{fig:chirality_check}
	\end{center}
\end{figure}

\end{document}